\documentclass[12pt,a4paper]{article}
\usepackage[T1]{fontenc}
\usepackage{fancyhdr}
\usepackage{amsmath,amsthm,amsfonts,amssymb}
\allowdisplaybreaks% page breaks allowed in align equations
\usepackage{epic,eepic}
\usepackage{graphicx}
\usepackage{authblk}
\usepackage{fullpage}
\usepackage[active]{srcltx}
\usepackage{enumitem}% itemize options
\usepackage{booktabs}% for tables

\usepackage{framed,color}

\usepackage{float}%
\floatstyle{plaintop}%
\restylefloat{table}% these three lines put caption for tables on top

\usepackage[font={small,it}]{caption}% the options makes the font italic in floats
\usepackage[table]{xcolor}% use colours in table rows

\usepackage[toc,page]{appendix}
\graphicspath{{Figures/}}

\usepackage[style=authoryear,backend=bibtex,maxcitenames=2,maxbibnames=99]{biblatex}
\bibliography{MDSBM}

\usepackage{chngcntr}
\usepackage{bbm}% indicator function \mathbbm{1} symbol

\usepackage{algpseudocode}
\usepackage{algorithm}
\usepackage{setspace}% for the pseudocode

\usepackage{tikz}
\usetikzlibrary{bayesnet}

\makeatletter
\AtEveryBibitem{%
  \global\undef\bbx@lasthash%
  \clearfield{}}
\makeatother

\theoremstyle{plain}

\theoremstyle{definition}

\theoremstyle{remark}

\numberwithin{equation}{section}
\numberwithin{figure}{section}

\date{\today}
\title{\textbf{Choosing the number of groups in a latent stochastic block model for dynamic networks}}
\author[1,2,*]{Riccardo Rastelli}
\author[3]{Pierre Latouche}
\author[1,2]{Nial Friel}
\affil[*]{\footnotesize riccardo.rastelli@ucdconnect.ie}
\affil[1]{\footnotesize School of Mathematics and Statistics, University College Dublin, Ireland;}
\affil[2]{\footnotesize Insight Centre for Data Analytics, Ireland;}
\affil[3]{\footnotesize Laboratoire SAMM, Universit\'e Paris 1 Panth\'eon-Sorbonne, France.}

\begin{document}
\rowcolors{2}{gray!25}{white}% use colours in table's rows
\counterwithout{figure}{section}
\counterwithout{figure}{subsection}
\counterwithout{equation}{section}
\counterwithout{equation}{subsection}

\maketitle
\begin{abstract}
\noindent
Latent stochastic block models are flexible statistical models that are widely used in social network analysis.
In recent years, efforts have been made to extend these models to temporal dynamic networks, whereby the connections between nodes are observed 
at a number of different times.
In this paper we extend the original stochastic block model by using a Markovian property to describe the evolution of nodes' cluster memberships over time.
We recast the problem of clustering the nodes of the network into a model-based context, 
and show that the integrated completed likelihood can be evaluated analytically for a number of likelihood models. 
Then, we propose a scalable greedy algorithm to maximise this quantity, thereby estimating both the optimal partition and the ideal number of groups 
in a single inferential framework.
Finally we propose applications of our methodology to both real and artificial datasets.
\\

\end{abstract}

\baselineskip=20pt
\section{Introduction}

In the  last few years,  there has been  an increasing amount  of data
stored characterising interactions  between  individuals or,  more  generally,
units of interest.  An interaction, represented by a triple $(i, j,
\eta)$,  indicates that  units $i$  and $j$  have a  connection at  a
specific time  point $\eta$.  Interactions can for instance
describe  email  exchanges  between  individuals or  posts  on  social
media. In Biology,  units can correspond to genes  and interactions to
regulation events between the genes.  A natural approach to
model the  set of all  observed interactions is  to rely on  a dynamic
graph where each unit is associated with  a node and an edge $(i, j)$ is
present  between  two units  $i$  and  $j$,  at  time $\eta$,  if  the
interaction $(i, j, \eta)$ is recorded. 

A long  series of methods have  been proposed recently to  cluster the
nodes of dynamic networks in order to summarise the information hidden
in such data. A large number of these  consider the Stochastic Block  Model (SBM)
\parencite{wang1987,nowicki2001estimation}  as   a  starting   point  and
propose extensions to the dynamic
framework.   Moreover,   they   are    usually   discrete   in   time,
that is, predefined time intervals  are introduced and interactions during
those time intervals  are aggregated to obtain snapshots  indexed by a
discrete time variable  $t$. In the case of a  binary dynamic network,
two  nodes are  connected in  a  snapshot if  they have  at least  one
interaction in the corresponding time interval. 
We recall that the SBM was originally introduced to cluster nodes
in  static  networks where  nodes  and  edges  do not  evolve  through
time. 
The model assumes that nodes are spread in latent clusters, which, in practice, have
to be  inferred from  the data.  The probability  for two
nodes  to  connect   is  then  only  dependent   on  their  respective
clusters.  Because  no  constraints  are  imposed  on  the  connection
probabilities, various  types of  clusters can  be extracted  from the
data, which makes methodologies based  on SBM applicable to  networks with different
connectivity structures \parencite{daudin2008mixture}.  The dynamic SBM-like model of
\textcite{yang2011detecting} allows, for example, each node to switch its cluster at
time $t$ depending on its state  at time $t-1$. A transition matrix is
employed to  characterise the switching probabilities.  While clusters
can   change   over   time,   fixed   connection   probabilities   are
used. Conversely, the model of \textcite{xu2014dynamic} relies on evolving
connection probabilities and  temporal changes are
described  through  a  state  space model.  Therefore,  the  inference
requires the use  of optimisation tools such as the  Kalman filter and
the Rauch-Tung-Striebel smoother. The work of \textcite{yang2011detecting}
was then  extended by \textcite{matias2015statistical} for  the clustering
of  nodes  in  dynamic  networks   where  edges  are  not  necessarily
binary. We emphasise that theoretical results are also provided in
their paper to show that dynamic SBM-like models should not
let both  the clusters and  connection parameters evolve  through time
without  incurring   into  identifiability   issues.  

In   the  static
framework,  many   extensions  have   been  considered  for   the  SBM
model.  Many of  them  have then  been adapted  to  deal with  dynamic
networks.  For  example,  the well-known mixed-membership SBM of \textcite{airoldi2009mixed}
has been extended to a dynamic framework by several works including \textcite{xing2010}, \textcite{dM3SBM} and \textcite{Kim2013}.
Note  that the latent position  model
of \textcite{hoff2002latent}, which is also popular in the network
community, was also adapted by \textcite{sarkar2005dynamic}
and \textcite{Friel2016} to deal with dynamic interactions. 

In  this paper,  our objective is to model the  evolution  of the  cluster
memberships over time  by relying on a Markovian  property.  
We take advantage of prior conjugacy to integrate out analytically most 
of the model parameters, with an approach similar to that of
\textcite{mcdaid2013improved} and \textcite{come2015model}.
This so-called collapsing allows one to obtain an analytical expression 
for the integrated completed data likelihood for 
a  number  of  likelihood   models.  A  greedy
optimisation algorithm is  then employed  for inferential  purposes. It  allows
estimation of both  the number of  clusters and the cluster  memberships of
the nodes to the clusters.  

Section \ref{Markovian  Stochastic Block Model}  introduces  the  model  and the notation.  A  Bayesian
hierarchical structure  is then proposed in Section
\ref{BayesianHierarchicalStructure}.  Finally, the optimisation procedure is
given in  Section \ref{sec:greedyopt} and experiments  are carried out
in Sections  \ref{sec:simu}, \ref{sec:enron}, and  \ref{sec:london} to
assess the proposed methodology. 

\section{Markovian Stochastic Block Model}\label{Markovian Stochastic Block Model}
The observed data consist of a sequence of network objects $\mathcal{X} = \left\{\textbf{X}^{(t)}\right\}_{t\in\mathcal{T}}$
defined on the same set of nodes $\mathcal{V} = \left\{ 1,\dots, N\right\}$. 
We focus our attention on dynamic networks, where $\mathcal{T} = \left\{ 1,\dots,T\right\}$ denotes the temporal span.

The random variable $X_{ij}^{(t)}$ models the value exhibited by the edge from $i$ to $j$ at time $t$, $\forall t \in \mathcal{T}$ and $\forall  i,\ j \in \mathcal{V}$.
We outline our methodology on directed networks, however it applies also to undirected network as we illustrate in the applications we consider. 
A typical scenario is that of a binary dynamic network, where:
\begin{equation}
\rowcolors{1}{}{}
 x_{ij}^{(t)} = \begin{cases}
                 1, & \mbox{ if an edge from $i$ to $j$ appears at time $t$; }\\
                 0, & \mbox{ otherwise. }
                \end{cases}
\end{equation}
Also, self-edges are not allowed, i.e. $x_{ii}^{(t)} = 0,\ \forall t\in\mathcal{T}$ and $\forall i \in\mathcal{V}$.

The random variable $Z_i^{(t)}$ characterises the allocation of a node, whereby $Z_i^{(t)} = g$ indicates that node $i$ is allocated to group $g$ at time $t$, 
for a certain $g \in \mathcal{K} = \left\{ 1,\dots,K\right\}$.
The set $\textbf{Z}^{(t)} = \left\{ Z_i^{(t)}\right\}_{i\in\mathcal{V}}$ therefore corresponds to a random hard clustering (i.e. a partitioning) of $\mathcal{V}$, $\forall t \in \mathcal{T}$.
We also denote the full set of allocations with $\mathcal{Z} = \left\{\textbf{Z}^{(t)}\right\}_{t\in\mathcal{T}}$.
The total number of groups in this underlying structure is denoted by $K$, hence $Z_i^{(t)} \in \mathcal{K},\ \forall t\in\mathcal{T}$ and $\forall i\in\mathcal{V}$.
Note that $K$ represents the total number of groups in $\mathcal{Z}$, implying that, at each time frame $t$, the current number of non-empty groups $K^{(t)}$ may be any number in $\mathcal{K}$.

The allocations characterise the connection profiles of the nodes of the network, in that nodes allocated to the same group have their edges drawn from the same probability distribution.
The edges of the dynamic network are marginally dependent, however they are conditionally independent given the allocations:
\begin{equation}
 p\left( \mathcal{X} \middle\vert \mathcal{Z} \right) = \prod_{t\in\mathcal{T}} p\left( \textbf{X}^{(t)}\middle\vert \textbf{Z}^{(t)} \right),
\end{equation}
and $\forall t \in \mathcal{T}$:
\begin{equation}
 p\left( \textbf{X}^{(t)}\middle\vert \textbf{Z}^{(t)} \right) = \prod_{i\in\mathcal{V}}\prod_{j\in\mathcal{V}: j\neq i} p\left( X_{ij}^{(t)}\middle\vert Z_i^{(t)}, Z_j^{(t)} \right).
\end{equation}
The distribution for the value of a single edge is determined by the allocations of the nodes, as follows:
\begin{equation}
 p\left( X_{ij}^{(t)} = x \middle\vert z_i^{(t)} = g, z_j^{(t)} = h, \boldsymbol{\Theta}\right) = f\left( x;\ \boldsymbol{\theta}_{gh} \right),
\end{equation}
where $\boldsymbol{\Theta}$ and $\boldsymbol{\theta}_{gh}$ are collections of model parameters.
In the binary network case, $f$ corresponds to the mass probability function of a Bernoulli variable with success probability $\theta_{gh}\in[0,1]$. 
In such case the likelihood of the model may be written as:
\begin{equation}
 \mathcal{L}_{\mathcal{X}} = \prod_{g\in\mathcal{K}}\prod_{h\in\mathcal{K}}\prod_{t\in\mathcal{T}}
 \prod_{\left\{i\in\mathcal{V}:\ z_i^{(t)}=g\right\}}\prod_{\left\{j\in\mathcal{V}:\ j\neq i;\ z_j^{(t)}=h\right\}}  \theta_{gh}^{x_{ij}^{(t)}}\left( 1-\theta_{gh} \right)^{1-x_{ij}^{(t)}}.
\end{equation}

Concerning the modelling of the temporal evolution of the network, a Markovian property on the nodes' allocations is adopted.
The sequences $\textbf{Z}_i = \left\{ Z_i^{(t)}\right\}_{t\in\mathcal{T}}$ are assumed to be independent Markov chains realised using the same $K\times K$ kernel matrix $\Pi$.
Therefore the prior structure on the allocations factorises as follows:
\begin{equation}\label{prior1}
 p\left( \mathcal{Z} \middle \vert \Pi\right) = 
 p\left( \textbf{Z}^{(1)} \right)\prod_{t\in\mathcal{T}\setminus\{1\}}p\left( \textbf{Z}^{(t)}\middle\vert \textbf{Z}^{(t-1)}, \Pi \right),
\end{equation}
where, $\forall t \in\mathcal{T}\setminus\{1\}$ and $\forall i\in\mathcal{V}$:
\begin{equation}\label{prior2}
 p\left( Z_{i}^{(t)} = h \middle\vert Z_{i}^{(t-1)} = g, \Pi\right) = \pi_{gh} \in [0,1].
\end{equation}
Note that the rows of $\Pi$ must sum to $1$, whereas no constraint is imposed on the columns of the same matrix.
Combining \eqref{prior1} and \eqref{prior2}, the prior on the allocations may be alternatively written as:
\begin{equation}\label{prior3}
 p\left( \mathcal{Z} \middle \vert \Pi\right) = 
 p\left( \textbf{Z}^{(1)} \right)\prod_{g\in\mathcal{K}}\prod_{h\in\mathcal{K}}\pi_{gh}^{R_{gh}},
\end{equation}
where $R_{gh}$ denotes the total number of switches from group $g$ to $h$.

\section{Bayesian hierarchical structure}\label{BayesianHierarchicalStructure}
We propose a hierarchical structure to further model the parameters $\boldsymbol{\theta}$ and $\Pi$.
In particular, we focus on the use of conjugate priors, since these allow one to integrate out (\textit{collapse}) most of the model parameters.

Concerning the transition probabilities, we specify a Dirichlet distribution over the rows of $\Pi$:
\begin{equation}
 \left( \pi_{g1},\dots,\pi_{gK} \right) \sim Dir\left( \delta_{g1},\dots,\delta_{gK} \right),\hspace{1cm}\forall g \in \mathcal{K}.
\end{equation}
By analytically integrating out the parameters $\left\{\pi_{gh}\right\}_{g,h}$, the following compound distribution (``marginal prior'') for the allocations arises:
\begin{equation}
 p\left( \mathcal{Z} \middle \vert \boldsymbol{\delta}\right) = p\left( \textbf{Z}^{(1)} \right)\prod_{g\in\mathcal{K}}
 \left\{ 
 \frac{\prod_{h\in\mathcal{K}}\Gamma\left( \delta_{gh}+R_{gh} \right)}{\Gamma\left( \sum_{h\in\mathcal{K}}\delta_{gh} + \sum_{h\in\mathcal{K}}R_{gh} \right)}
 \frac{\Gamma\left( \sum_{h\in\mathcal{K}}\delta_{gh} \right)}{\prod_{h\in\mathcal{K}}\Gamma\left( \delta_{gh} \right)}
 \right\}.
\end{equation}
We assume a uninformative flat prior distribution for the hyperparameters $\left\{\delta_{gh}\right\}_{g,h}$ and thus fix all of them to $1$.

A separate and independent Multinomial-Dirichlet model may be specified for $\textbf{Z}^{(1)}$. However, for these starting allocations, 
we opt for a more pragmatic and parsimonius approach: we use a Multinomial distribution with parameters $\left\{ \alpha_g\right\}_{g\in\mathcal{K}}$, where:
\begin{equation}
 \alpha_g \propto \sum_{t\in\mathcal{T}\setminus\left\{1\right\}} \sum_{i\in\mathcal{V}} \mathbbm{1}_{\left\{ Z_i^{(t)} = g\right\}}, \hspace{1cm}\forall g\in\mathcal{K}.
\end{equation}
The distribution on the initial states so defined approximates the stationary distribution associated to $\Pi$.

Regarding the likelihood structure, a number of conjugate models can be considered, encompassing most edges' types. 
Table \ref{tab:1} provides a list of possible scenarios that may be of interest.
\begin{table}[ht]
\caption{A list of edge types that can be accounted for using the methodology proposed. For each case the corresponding Bayesian hierarchical structure is shown.} 
\label{tab:1}
\centering
\begin{tabular}{ccc}
  \toprule
  Edge type & Likelihood & Conjugate prior \\ 
  \midrule
  Binary & Bernoulli & Beta \\
  Categorical & Multinomial & Dirichlet \\
  Counts (positive integers) & Poisson & Gamma \\
  Positive weights & Gamma & Gamma \\
  Truncated counts or proportions & Binomial & Beta \\
  Heavy tailed positive weights & Pareto & Gamma \\
  Real numbers & Normal & Normal-Gamma \\
  Real numbers with covariates & Multivariate Normal & Normal-Gamma\\
  \bottomrule
\end{tabular}
\end{table}
In this paper, we focus on binary edges, and hence specify an independent Beta prior on the likelihood parameters 
$\Theta = \left\{ \theta_{gh}:\ g\in \mathcal{K},\ h\in \mathcal{K}\right\}$.
In such a case the compound distribution for $\mathcal{X}$ (``marginal likelihood'') is:
\begin{equation}
 p\left( \mathcal{X} \middle\vert \mathcal{Z}, a, b\right) = \prod_{g\in\mathcal{K}}\prod_{h\in\mathcal{K}}
 \left\{ \frac{\Gamma\left( a+b \right)}{\Gamma\left( a \right)\Gamma\left( b \right)} 
 \frac{\Gamma\left( a + \eta_{gh}\right) \Gamma\left( b+N_{gh}-\eta_{gh} \right)}{\Gamma\left( a+b+N_{gh} \right)}\right\},
\end{equation}
where $a$ and $b$ are the Beta hyperparameters, $\eta_{gh}$ counts the number of edges from group $g$ to $h$:
\begin{equation}
 \eta_{gh} = \sum_{t\in\mathcal{T}}\sum_{\left\{i\in\mathcal{V}:\ z_i^{(t)}=g\right\}}\sum_{\left\{j\in\mathcal{V}:\ j\neq i;\ z_j^{(t)}=h\right\}} X_{ij}^{(t)}
\end{equation}
and $N_{gh}$ counts the total number of possible edges from group $g$ to $h$:
\begin{equation}
N_{gh} = \sum_{t\in\mathcal{T}}\sum_{\left\{i\in\mathcal{V}:\ z_i^{(t)}=g\right\}}\sum_{\left\{j\in\mathcal{V}:\ j\neq i;\ z_j^{(t)}=h\right\}} 1.
\end{equation}
Note that, similarly to the hyperparameter $\delta$, we make a ``symmetric'' assumption on the hyperparameters $a$ and $b$, making them identical over all of the groups.
In the applications, these are both fixed to $1$ yielding a uninformative flat prior distribution.

\begin{figure}[htbp]
 \centering
\begin{tikzpicture}[->,>=stealth',shorten >=1pt,auto,node distance=3cm,thick,main node/.style={circle,draw,font=\sffamily\Large\bfseries}]
  \node[obs, shape=rectangle, xshift=1.7cm, fill=white, minimum size=1cm]  (delta) at (1,1.5) {\LARGE{$\boldsymbol{\delta}$}};
  \node[obs, shape=rectangle, xshift=1.7cm, fill=white, minimum size=1cm]  (phi) at (14,1.5) {\LARGE{$\boldsymbol{\phi}$}};
  \node[obs, shape=rectangle, xshift=1.7cm, fill=white, minimum size=1cm]  (pi) at (3,1.5) {\LARGE{$\boldsymbol{\pi}$}};
  \node[obs, shape=rectangle, xshift=1.7cm, fill=white, minimum size=1cm]  (theta) at (12,1.5) {\LARGE{$\boldsymbol{\theta}$}};
  \node[obs, shape=rectangle, xshift=1.7cm, fill=white, minimum size=1.75cm]  (Ztm1) at (6,0) {\LARGE{$\textbf{Z}$}\textsuperscript{\footnotesize $(t-1)$}};
  \node[obs, shape=rectangle, xshift=1.7cm, fill=white, minimum size=1.75cm]  (Zt) at (6,3) {\LARGE{$\textbf{Z}$}\textsuperscript{\footnotesize $(t)$}};
  \node[obs, shape=circle, xshift=1.7cm, fill=white, minimum size=2cm]  (Xtm1) at (9,0) {\LARGE{$\textbf{X}$}\textsuperscript{\footnotesize $(t-1)$}};
  \node[obs, shape=circle, xshift=1.7cm, fill=white, minimum size=2cm]  (Xt) at (9,3) {\LARGE{$\textbf{X}$}\textsuperscript{\footnotesize $(t)$}};
%   Invisible node for white space
  \node[] (invtop) at (7.7,5) {};
  \node[] (invbot) at (7.7,-2) {};
  % edges
  \draw (delta) -- (pi) node [midway, fill=white, scale=0.5, above=-0.35] {};
  \draw (phi) -- (theta) node [midway, fill=white, scale=0.5, above=-0.35] {};
  \draw (theta) -- (Xtm1) node [midway, fill=white, scale=0.5, above=-0.35] {};
  \draw (theta) -- (Xt) node [midway, fill=white, scale=0.5, above=-0.35] {};
  \draw (pi) -- (Ztm1) node [midway, fill=white, scale=0.5, above=-0.35] {};
  \draw (pi) -- (Zt) node [midway, fill=white, scale=0.5, above=-0.35] {};
  \draw (Ztm1) -- (Zt) node [midway, fill=white, scale=0, above=-0.35] {};
  \draw (Ztm1) -- (Xtm1) node [midway, fill=white, scale=0, above=-0.35] {};
  \draw (Zt) -- (Xt) node [midway, fill=white, scale=0, above=-0.35] {};
  \draw [-,dashed] (Zt) to (invtop);
  \draw [-,dashed] (invbot) to (Ztm1);
\end{tikzpicture}
\caption{Graphical model for the Markovian stochastic block model described.}
\label{fig:graphical_model}
\end{figure}
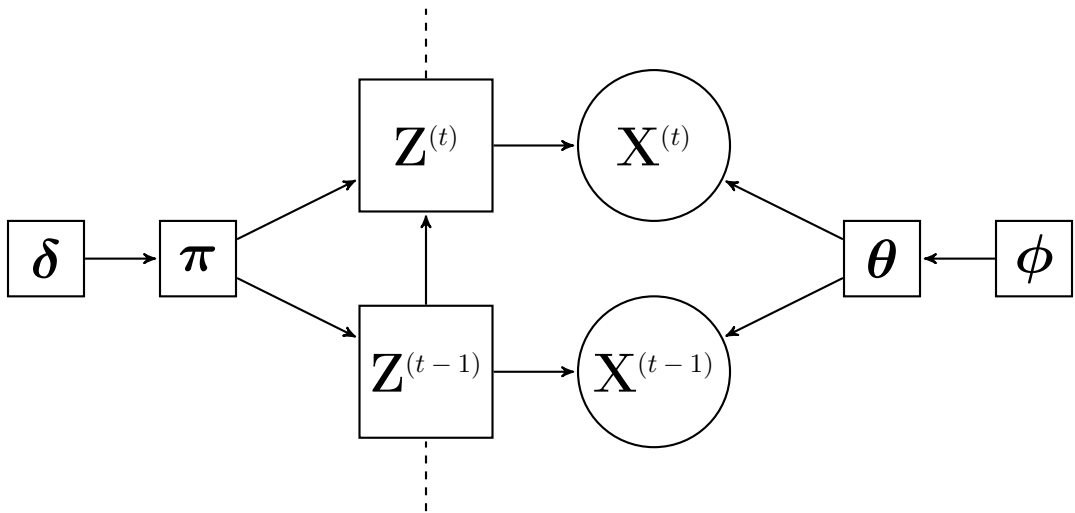

Figure \ref{fig:graphical_model} shows a graphical model summarising the hierarchical structure and the dependencies between the variables introduced.

A similar modelling framework has been recently proposed by \textcite{matias2015statistical}.
In this paper, we take advantage of some of their results on model-identifiability and we propose a radically different estimation technique.

\subsection{Incomplete weighted networks}\label{Handling absent edges in weighted networks}
When dealing with weighted networks, graphs are often incomplete, in that not all the edges exhibit a value. 
The presence-absence of edge values can be modelled using a Bernoulli random variable $\rho_{ij}$, which is equal to one if the edge from $i$ to $j$ is present and is $0$ otherwise.
This adds an extra layer in the modelling, which is unrelated to the likelihood model specified on the present edge values.
In fact, any likelihood structure can be used to model the edges that appear ($\rho_{ij}=1$) in the network.

Assuming a Beta prior on the presence-absence indicator, the marginal likelihood of the network is simply the product of the two compound distributions:
\begin{equation}
 p\left( \mathcal{X}\middle\vert \mathcal{Z}, a_{\rho}, b_{\rho}, \boldsymbol{\phi} \right) = 
 p\left( \boldsymbol{\rho}\middle\vert \mathcal{Z}, a_{\rho}, b_{\rho}\right) p\left( \mathcal{X}\middle\vert \mathcal{Z}, \boldsymbol{\rho}, \boldsymbol{\phi} \right);
\end{equation}
where $\left(a_{\rho}, b_{\rho}\right)$ and $\boldsymbol{\phi}$ denote the hyperparameters for $\boldsymbol{\rho}$ and $\boldsymbol{\theta}$, respectively.
Note that the modelling on $\boldsymbol{\rho}$ can capture the heterogeneity induced by the block structure, in that the probability of an edge exhibiting a value may depend on its allocation.

\section{Exact Integrated Completed Likelihood}\label{ExactIntegratedCompletedLikelihood}
In the previous section, we have shown that in a binary Markovian dynamic network the marginal likelihood $p\left( \mathcal{X}\middle\vert\mathcal{Z},a,b \right)$ 
and the marginal prior $p\left( \mathcal{Z}\middle\vert \delta\right)$ have an exact analytical form. 
% The same result can be achieved in a number of likelihood frameworks, as shown in Table \ref{tab:1}.
These terms can be recombined to obtain a particularly meaningful quantity, the so-called exact Integrated Completed Likelihood (ICL), defined as:
\begin{equation}\label{icl1}
 \mathcal{ICL}_{ex} = \log\left[ p\left( \mathcal{X}\middle\vert\mathcal{Z},a,b \right)\right] + \log\left[ p\left( \mathcal{Z}\middle\vert \delta\right) \right].
\end{equation}
Such a quantity corresponds to the exact value that the ICL criterion of \textcite{biernacki2000assessing} propose to maximise when choosing the number of groups in a finite mixture context.
We stress that, although the definition in \eqref{icl1} relates to the specific case of binary dynamic networks, the same quantity can be evaluated 
analytically for all of the likelihood models listed in Table \ref{tab:1}.

We propose to use $\mathcal{ICL}_{ex}$ as a model-based optimality criterion for the clustering problem on the nodes of our dynamic network: in the space of all possible allocations, 
we seek a $\hat{\mathcal{Z}}$ maximising $\mathcal{ICL}_{ex}$.
Note that, thanks to the discrete nature of the allocation variables, the optimal total number of groups $\hat{K}$ can be deduced automatically from $\hat{\mathcal{Z}}$, along with the values $\left\{K^{(t)}\right\}_{t\in\mathcal{T}}$.

We also emphasise that, from a Bayesian perspective, $\hat{\mathcal{Z}}$ corresponds to a Maximum A Posteriori (MAP) solution:
\begin{equation}
 p\left( \mathcal{X}\middle\vert\mathcal{Z},a,b \right)p\left( \mathcal{Z}\middle\vert \delta\right) = p\left( \mathcal{X},\mathcal{Z}\middle\vert a,b,\delta \right) 
 \propto p\left( \mathcal{Z}\middle\vert \mathcal{X}, a,b,\delta \right),
\end{equation}
where the proportionality is intended with respect to $\mathcal{Z}$.

\section{Greedy optimisation}\label{sec:greedyopt}
In the Markovian dynamic stochastic block model described, we seek a clustering solution $\mathcal{\hat{Z}}$ maximising the $\mathcal{ICL}_{ex}$ value defined in \eqref{icl1}.
A complete exploration of all possible allocations is not feasible, even for small datasets. 
However, similar combinatorial problems have been recently tackled successfully using heuristic greedy routines, resembling the well known Iterated Conditional Modes of \textcite{besag1986statistical}. 
Greedy algorithms have been applied in a hierarchical clustering framework for networks by \cite{newman2004fast}. 
More recently, they have been adapted to maximise the $\mathcal{ICL}_{ex}$ in a model-based clustering context for networks in \textcite{come2015model,wyse2014inferring} and \textcite{corneli2016exact} 
and for Gaussian finite mixtures in \textcite{bertoletti2015choosing}.
Here, we propose an extension of the same ideas to our dynamic network context. 

The algorithm repeatedly sweeps over the network's nodes and attempts to reallocate them using a greedy behaviour. 
In each step, a random time frame $t$ and a random node $i$ are chosen, and the variation of $\mathcal{ICL}_{ex}$ is evaluated for all the possible reallocations of the corresponding node. 
For every group $g$, the value $\ell_{\left( t,i \right)\rightarrow g}$ is evaluated, corresponding to the $\mathcal{ICL}_{ex}$ value after moving the node identified by $\left( t,i \right)$ to $g$.
Eventually the node is reallocated to the group that yields the best increase in the objective function. This procedure continues until no reallocation can provide a further increase.
The corresponding pseudocode is shown in Algorithm \ref{GreedyICL}. 
\begin{algorithm}[htb]
\begin{spacing}{1.2}
\caption{\texttt{GreedyIcl}}
\label{GreedyICL}
\begin{algorithmic}
\State Initialise $Z_i^{(t)}$, $\forall t\in\mathcal{T}$, $\forall i\in\mathcal{V}$.
\State Evaluate $\mathcal{ICL}_{ex}$ and set $\ell = \mathcal{ICL}_{ex}$ and $\ell_{stop} = \mathcal{ICL}_{ex}$.
\State Set $stop = false$.
\While{$!stop$}
\State Set $\mathcal{U} = \left\{ \left( t,i\right):\ t\in\mathcal{T},\ i\in \mathcal{V}\right\}$.
\State Shuffle the elements of $\mathcal{U}$.
\While{$\mathcal{U}$ is not empty}
\State $\left( t,i \right) = pop\left( \mathcal{U} \right)$.
\State $\hat{g}=\arg \max_{\substack{g=1,2,\dots,K_{up}}} \ell_{\left( t,i \right)\rightarrow g}$.
\State $\ell=\ell_{\left( t,i \right)\rightarrow \hat{g}}$.
\State $Z_i^{(t)} = \hat{g}$.
\EndWhile
\If{$\ell \leq \ell_{stop}$} $stop = true$ \textbf{else} $\ell_{stop} = \ell$.
\EndIf
\EndWhile
\State Return $\mathcal{Z}$ and $\ell$.
\end{algorithmic}
\end{spacing}
\end{algorithm}

As input, the algorithm requires a starting partition, the hyperparameters' values, and a parameter $K_{up}$ denoting the largest admissible number of groups. 
During the optimisation, groups may be deleted (if they remain empty at all time frames) and created (if a node is reallocated to an empty group). 
In practice, the former case is very frequent whereas the latter is rather rare. 
Hence the best performance is achieved when the starting partition is composed of $K_{up}$ groups, with $K_{up}$ set to a fairly large value. 

The algorithm is bound to find only a local optimum for the objective function, and its initialisation plays a crucial role. 
Due to the random update order, it is possible that different starting partitions return different solutions, however we find that good initialisations often lead to a unique maximum.

As concerns the complexity of the algorithm, in the binary case one iteration can be performed in $\mathcal{O}\left( M+TNK_{up}^2 \right)$, where $M$ is the total number of edges appearing in the network.
Once a node $\left( t,i \right)$ is selected, the number of edges to and from group $g$ are counted, for every $g\in\left\{1,\dots,K_{up}\right\}$.
Evidently this implies a cost of $\mathcal{O}\left( m \right)$, where $m$ is the average value for the sum of in-degree and out-degree of a random node.
Then, the quantities $\ell_{\left( t,i \right)\rightarrow g}$ are evaluated for every $g$. 
For each of these, the computational bottleneck is given by the calculation of the variation of the marginal likelihood value, which can be performed in $\mathcal{O}\left( K_{up} \right)$.
This makes the average cost of updating an allocation $\mathcal{O}\left( m + K_{up}^2 \right)$.
Since all of the allocations are repeatedly updated, the overall complexity of one iteration is $\mathcal{O}\left( TNm+TNK_{up}^2 \right)$ as previously claimed.

\subsection{Final merge step}
Once the \texttt{GreedyIcl} has converged, we additionally propose a hierarchical clustering on the optimal solution, following the approach of \textcite{come2015model}. 
We consider all possible pairs of groups and attempt to merge each pair into a single cluster. 
Such a move is accepted only if the $\mathcal{ICL}_{ex}$ value increases, and everytime a merge move is performed the procedure is restarted.
For each try, the computational bottleneck is given by the evaluation of the variation for the marginal prior, which can be performed in $\mathcal{O}\left( \hat{K}^2 \right)$.
Since the number of pairs is $\mathcal{O}\left( \hat{K}^2 \right)$, and the number of accepted merge moves is capped at $\hat{K}$, the overall complexity of this final step is $\mathcal{O}\left( \hat{K}^5 \right)$.
Here $\hat{K}$ denotes the number of groups for the optimal solution obtained through the \texttt{GreedyIcl}, which is normally much smaller than $K_{up}$.
We find that in practice this final merge step does not impact the overall computational time by much, and the additional computational burden may be neglected.

\section{Simulation study}\label{sec:simu}
We propose an experiment to validate our methodology and compare it to the existing algorithm of \textcite{matias2015statistical}, using artificial data.
The number of time frames, the number of nodes, and the true underlying number of groups are fixed throughout as follows: $T = 4$, $N = 50$ and $K = 4$, respectively.
As concerns the transition probabilities matrices, the following general structure is assumed:
\begin{equation}\rowcolors{1}{}{}
 \Pi = \left(\begin{matrix}
  \pi & \nu & \nu & \nu \\
  \nu & \pi & \nu & \nu \\
  \nu & \nu & \pi & \nu \\
  \nu & \nu & \nu & \pi\\
 \end{matrix}\right)
\end{equation}
where $\nu = (1-\pi)/(K-1)$, so that rows sum to one. Choosing a high $\pi$ value, the clusters tend to be very stable and allocations do not change often over time. 
By contrast, a small $\pi$ value represents the opposite scenario, corresponding to a highly instable system. 
In either case, the stationary distribution, which is used to generate the starting partition, is $\boldsymbol{\alpha} = \left\{\frac{1}{K},\frac{1}{K},\frac{1}{K},\frac{1}{K}\right\}$.
In our simulations we consider two different scenarios: $\pi = 0.7$ and $\pi = 0.9$, which are denoted \textit{low-stability} and \textit{high-stability}, respectively. 
Although these two cases may offer only a limited view of all the situations encompassed by the model, we believe these to be the most interesting examples and that most realised networks are in fact well captured by this representation.

As concerns the connection probability matrices, an affiliation structure is assumed:
\begin{equation}\rowcolors{1}{}{}
 \left(\begin{matrix}
  \theta_0 & \epsilon_0 & \epsilon_0 & \epsilon_0 \\
  \epsilon_0 & \theta_0 & \epsilon_0 & \epsilon_0 \\
  \epsilon_0 & \epsilon_0 & \theta_0 & \epsilon_0 \\
  \epsilon_0 & \epsilon_0 & \epsilon_0 & \theta_0 \\
 \end{matrix}\right).
\end{equation}
A small perturbation is added independently to each of the entries of such matrix:
\begin{equation}\label{epsilon1}
 \theta = \theta_0 + 0.1u\hspace{2cm}\epsilon = \epsilon_0 + 0.1u,
\end{equation}
where $u$ is drawn from a Uniform distribution in the interval $[-1,1]$ for each entry. 
While $\epsilon_0$ is always fixed to $0.1$, $\theta_0$ is different for each matrix and determines the difficulty level in recovering the underlying latent clustering.
We consider $9$ different scenarios, each corresponding to a different choice of $\theta_0$, ranging from $0.1$ to $0.9$.
The perturbation created by the random variable $u$ is necessary to ensure identifiability of the model, as explained in \textcite{matias2015statistical}.\\

The experiment is executed as follows: for every choice of $\pi$ and $\theta_0$, $100$ random dynamic undirected networks are generated, each corresponding to a different random realisation of the connection probability matrix.
In each network, the variational algorithm of \textcite{matias2015statistical}, implemented in the \texttt{R} package \texttt{dynsbm} (version \texttt{0.3}), and our implementation of the greedy algorithm are executed.
As concerns the variational approach, \texttt{dynsbm} is run $4$ times for every value of $K$ in the set $\left\{1,2,\dots,6\right\}$, 
and the best run is retained as optimal solution according to the approximate ICL criterion advocated in \textcite{matias2015statistical}. 
Higher values of $K$ make \texttt{dynsbm} particularly time consuming and hence are not considered, however, we note that in all of the runs the approximate ICL criterion never favours the model with $6$ groups.
As concerns our greedy algorithm (denoted \texttt{GreedyIcl}), we consider several types of initialisations, as follows:
\begin{itemize}
 \item \texttt{aggregated}: an adjacency matrix of size $N\times N$ is obtained, by aggregating (summing) the $T$ adjacency matrices of the generated network. 
 Then, the kmeans algorithm is run on such a matrix using as number of centres a random draw from the discrete interval 
 $\lfloor 0.5*N\rfloor, \dots, \lfloor 0.75*N\rfloor$.
 This associates a cluster membership to each of the $N$ nodes. 
 These allocations are assumed not to change over time, so that the initial partition is given by the output of the kmeans repeated over all the time frames.
 \item \texttt{colbind}: the only difference with the \texttt{aggregated} initialisation lies in the fact that kmeans is run on a matrix obtained by gathering the adjacency matrices one next to the other, obtaining a $N \times TN$ matrix.
 Note that this is the same initialisation used for the \texttt{dynsbm} algorithm.
 \item \texttt{rowbind}: in this case kmeans is instead run on a matrix obtained by stacking up the observed adjacency matrices. 
 Since the size of this matrix is $TN\times N$, the number of groups considered by kmeans is not capped at $N$ as in the other cases: 
 hence we use a draw from the discrete interval $\lfloor 0.5*T*N\rfloor, \dots, \lfloor 0.75*T*N\rfloor$ as number of centres.
 Also, in contrast with the previous cases, kmeans returns the allocations for every node at every time frame, hence allocations are not assumed to be unchanged over time.
 \item \texttt{random}: the \texttt{GreedyIcl} is initialised unsing a random partition, with $K_{up}$ chosen as in \texttt{GreedyIcl rowbind}.
\end{itemize}
Once initialised, the \texttt{GreedyIcl} algorithm is run once for each type of initialisation.
% Then, for each run, only the overall best $\mathcal{ICL}_{ex}$ value and corresponding allocations are retained as optimal solutions.
The additional label 
\begin{itemize}
 \item \texttt{all}: is used to indicate the best solution obtained through \texttt{GreedyIcl} over all possible types of initialisations.
\end{itemize}

The optimal clusterings obtained using each method are compared to the true allocations using the \textit{Normalised Mutual Information} (NMI) index \parencite{strehl2003cluster}. 
This index takes values in the interval $[0,1]$ and describes how similar two partitions are. 
These partitions must be vectors, hence the optimal clusterings obtained are vectorised by concatenating the partitions time-wise.

\begin{figure}
 \centering
 \includegraphics[width=0.49\textwidth,page=1]{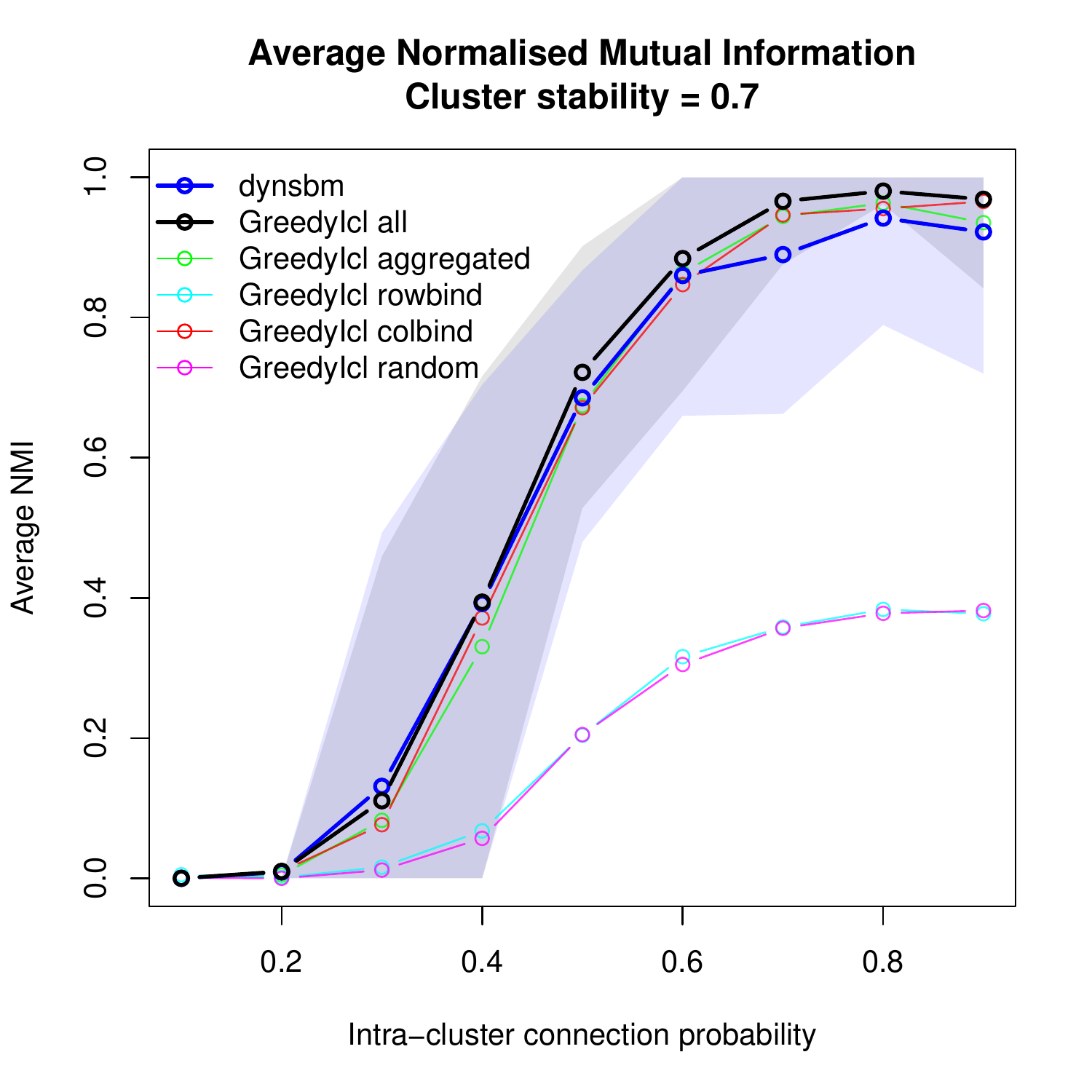}
 \includegraphics[width=0.49\textwidth,page=2]{simulation_study_nmi_mean.pdf}
 \caption{\textbf{Simulation study}. Average Normalised Mutual Information index between the true clustering and the optimal clustering solutions obtained through \texttt{dynsbm} (version \texttt{0.3}) and several versions of \texttt{GreedyIcl}.
 Shaded regions represent the $90\%$ quantile region associated to each scenario. These regions are plotted only for \texttt{GreedyIcl all} (black colour) and \texttt{dynsbm} (blue colour).
 It appears that the greedy algorithm with a good initialisation performs as well as the existing algorithm \texttt{dynsbm}.}
 \label{fig:simulation_study_nmi_mean}
\end{figure}

Figure \ref{fig:simulation_study_nmi_mean} shows the results obtained regarding the clustering performance.
It appears that the greedy optimisation coupled with kmeans initialisations and \texttt{dynsbm} achieve the best results, whereas the greedy optimisation with random intialisations perform poorly.
The average computing times are provided in Table \ref{tab:simulation_study_computing_time_1}.

\begin{table}
\centering
\caption{\textbf{Simulation study}. Average computing time required by each of the algorithms to be run on a generated dynamic network. 
Note that the variational expectation-maximisation of \texttt{dynsbm} and the greedy updates of our algorithm are both run exactly $4$ times for each dataset.
The initialisation may be different, but we find that this does not impact the computing time by much.}
\label{tab:simulation_study_computing_time_1}
\begin{tabular}{lc}    \toprule
Algorithm & Average computing time (seconds)  \\\midrule
\texttt{dynsbm} & 35.46  \\ 
\texttt{GreedyIcl all} & 4.58\\
\texttt{GreedyIcl aggregated} & 1.18  \\ 
\texttt{GreedyIcl colbind} & 1.19  \\ 
\texttt{GreedyIcl rowbind} & 1.14  \\ 
\texttt{GreedyIcl random} & 1.07  \\ \bottomrule
\hline
\end{tabular}
\end{table}

\begin{figure}
 \centering
 \includegraphics[width=0.49\textwidth,page=1]{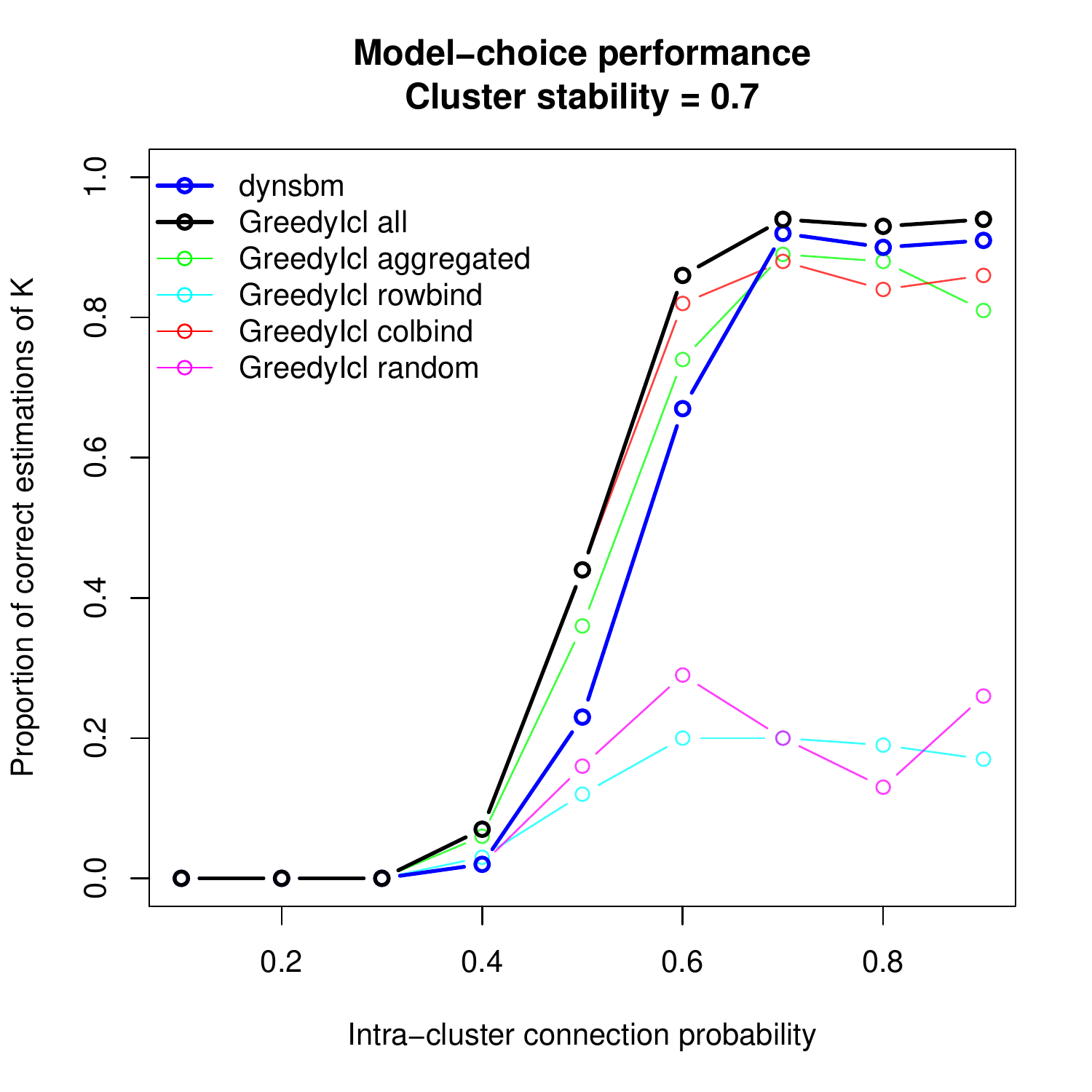}
 \includegraphics[width=0.49\textwidth,page=2]{simulation_study_k.pdf}
 \caption{\textbf{Simulation study}. For each combination of $\pi$ and $\theta_0$, the proportion of networks where $K$ is properly estimated is shown, for all of the algorithms considered.}
 \label{fig:simulation_study_k}
\end{figure}

As concerns model selection, our methodology outperforms the existing algorithm of \cite{matias2015statistical}, as shown in Figure \ref{fig:simulation_study_k}.

\section{Enron dataset}\label{sec:enron}
\paragraph{The data.}
The Enron Corporation filed for bankruptcy in late 2001, leading to an unprecedented scandal and to dire consequences for the US stock market. 
The data we use in this paper consists of all the emails exchanged from January 2000 to March 2002 between the Enron members.
This data was originally made public, and posted to the web, by the Federal Energy Regulatory Commission during its investigation.

\begin{figure}
 \centering
 \includegraphics[width=0.49\textwidth]{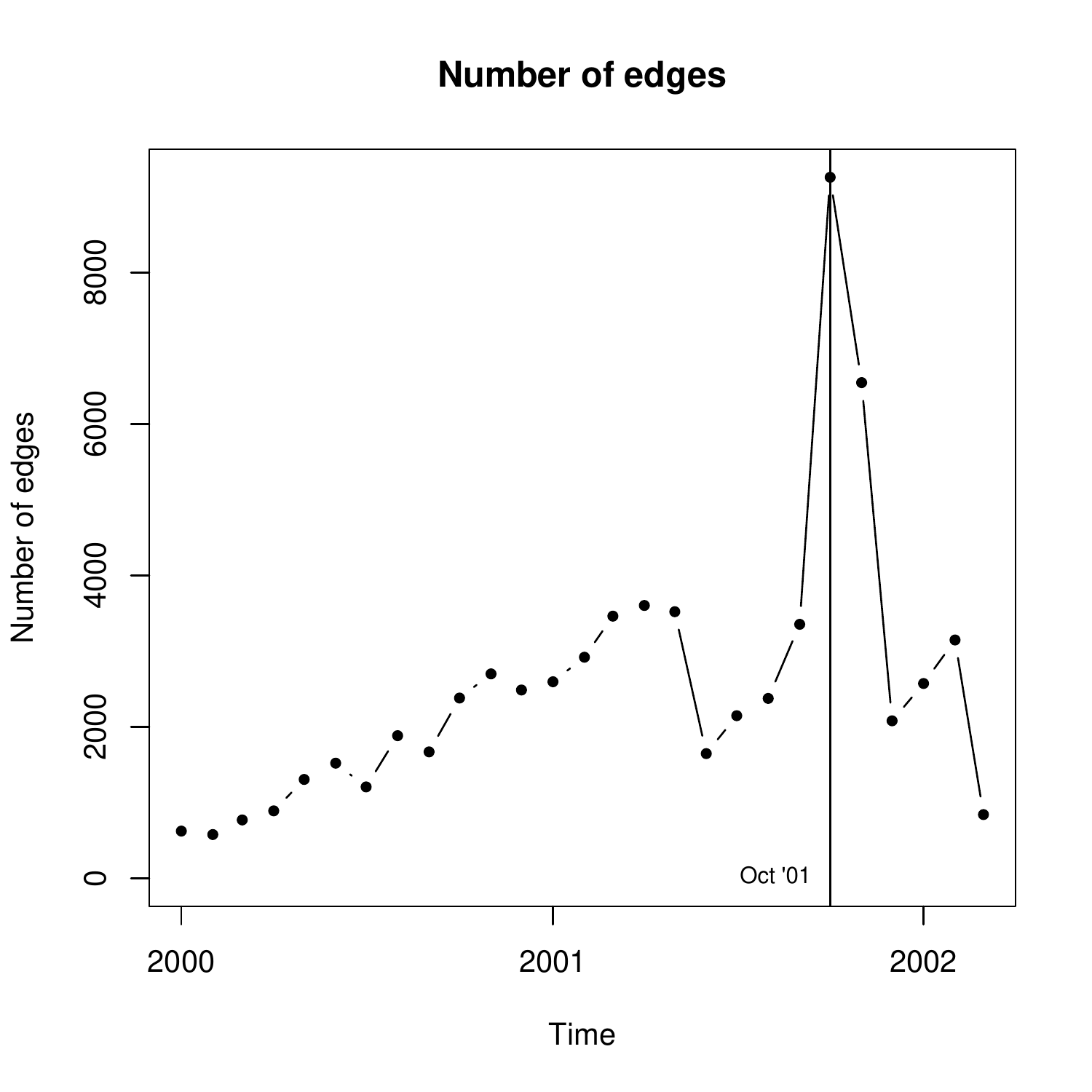}
 \caption{Number of edges at each time frame for the Enron email dataset. The peak in October 2001 corresponds to the disclosure of bankruptcy.}
 \label{fig:enron_n_edges}
\end{figure}

We construct a binary directed dynamic network of emails transforming the data into an adjacency cube $\mathcal{X}$, such that:
\begin{equation}\rowcolors{2}{gray!25}{white}% use colours in table's rows
x_{ij}^{(t)} = 
 \begin{cases}
  1, & \mbox{ at least an email is sent from member $i$ to member $j$ at time frame $t$;}\\
  0, & \mbox{ otherwise.}\\
 \end{cases}
\end{equation}
with each time frame $t$ corresponding to a one month period. The number of nodes is $N=148$ and the number of time frames is $T = 27$.
The number of edges observed at each time is shown in Figure \ref{fig:enron_n_edges}.
Along with the email data, the status of each member within the company is known, and is one of the following:
\textit{CEO}, \textit{Director}, \textit{Employee}, \textit{In House Lawyer}, \textit{Manager}, \textit{Managing Director}, \textit{N/A}, \textit{President}, \textit{Trader}, \textit{Vice President}.
To simplify the exposition of the results, we gather every status other than \textit{Employee} and \textit{N/A} into a single class named \textit{Manager}, so that only three status classes are considered.

We run our \texttt{GreedyIcl all} algorithm as in the simulation study, with each single run of the algorithm taking on average about $400$ seconds.
The overall best solution has $17$ groups. An analysis of this clustering solution follows.

\begin{figure}
 \centering
 \includegraphics[width=0.49\textwidth]{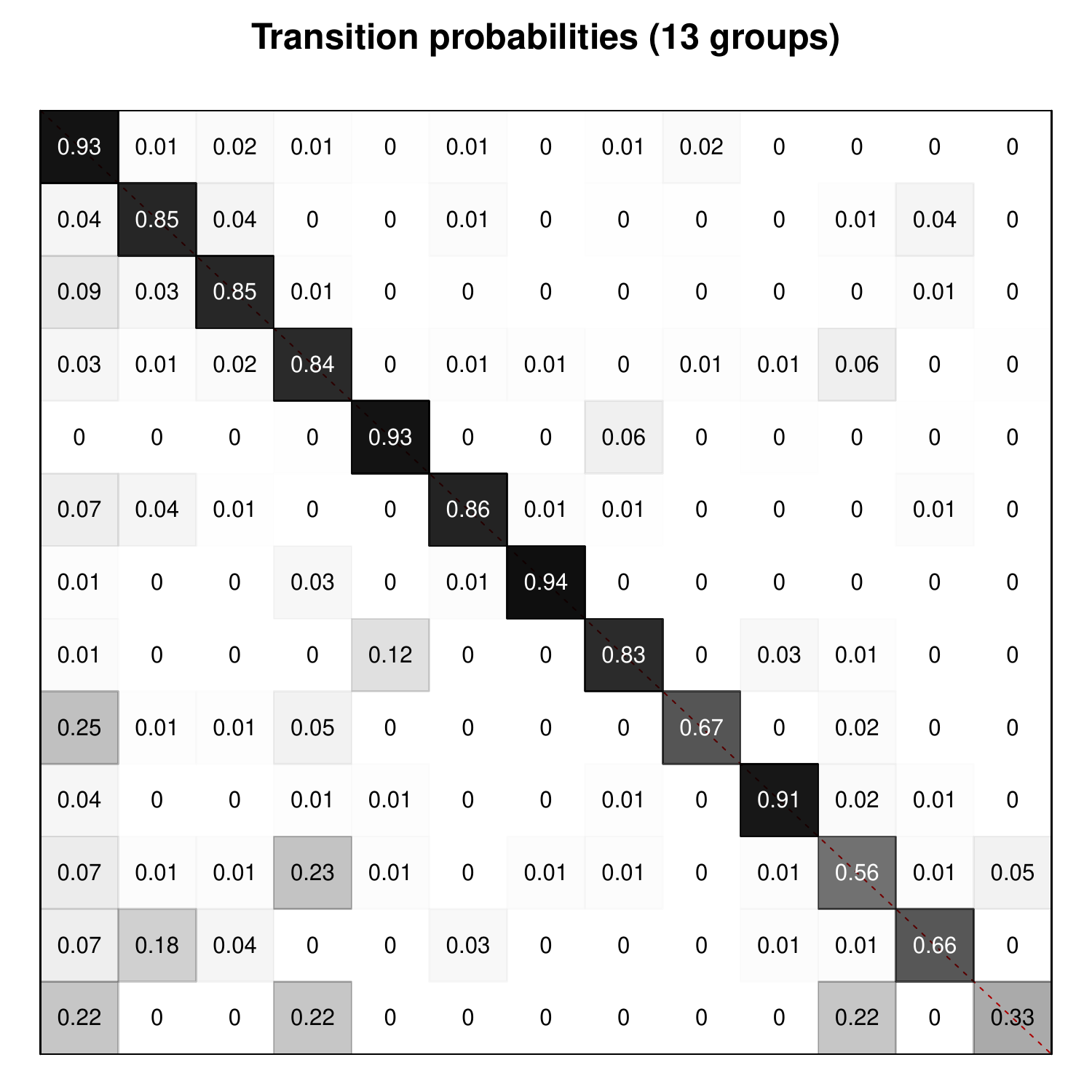}
 \includegraphics[width=0.49\textwidth]{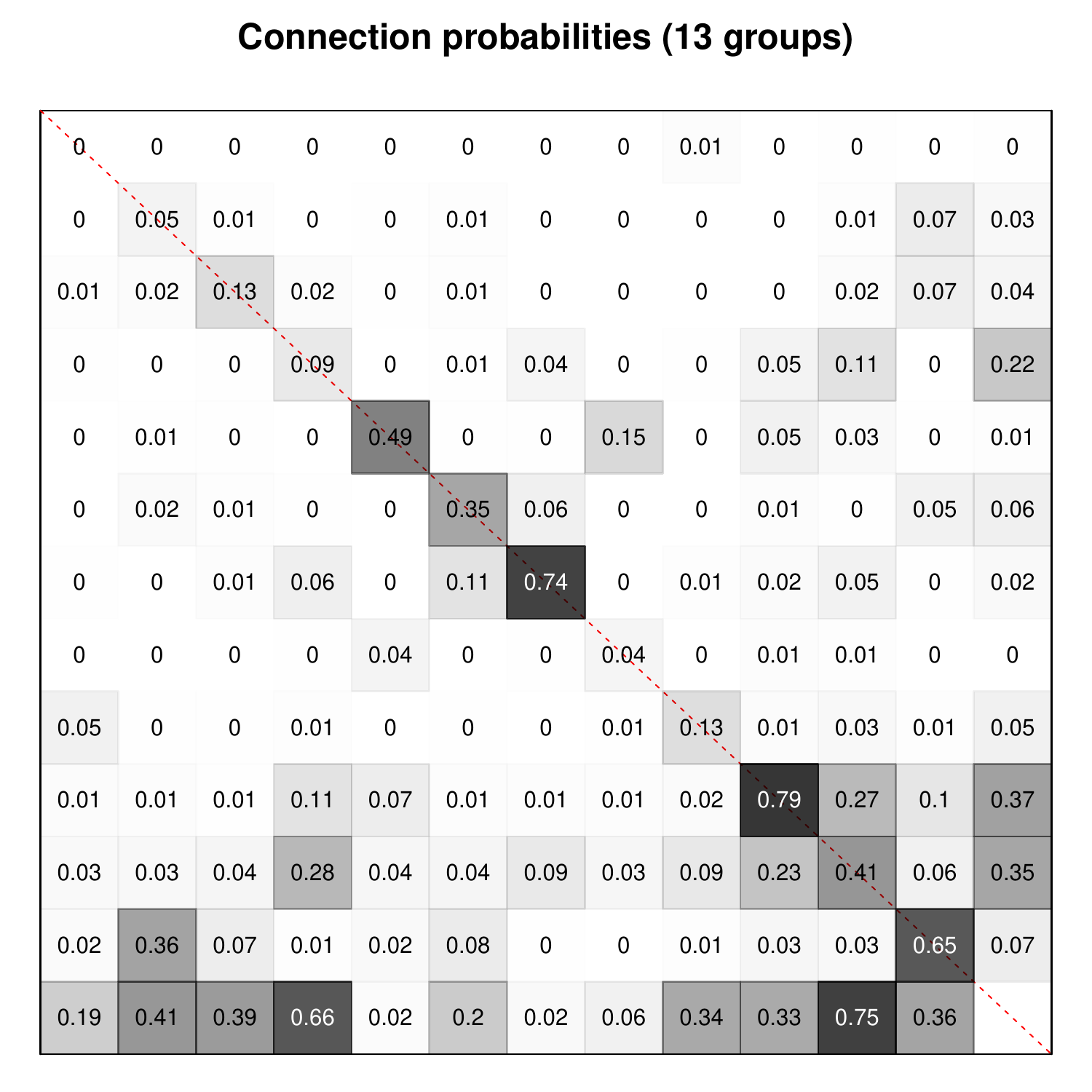}
 \caption{\textbf{Enron dataset}. On the left panel, the estimated transition probabilities are shown. It appears that the network is very stable over time, and nodes tend not to change their allocations often.
 On the right panel, the estimated connection probabilities are shown. Most groups have high within cluster connectivity, although a number of off-diagonal entries are also relevant. 
 The groups are ordered according to their aggregated size (i.e. the sum of their sizes over all times), in descending order. 
%  Hence the value in position $(1,1)$ shows the connection probability between two nodes belonging to the same largest group.
 Some entries cannot be properly estimated due to a small cluster size, hence are left blank.}
 \label{fig:enron_connectivity}
\end{figure}

\paragraph{Activity levels and connectivity.}
As shown in the left panel of Figure \ref{fig:enron_connectivity}, the clustering solution has very high stability over time, in that nodes do not change allocations frequently.
The connection probability matrix, shown on the right panel, exhibits instead a much more complex situation. 
Evidently, a strong community structure is present, since most of the diagonal elements of this matrix are fairly large.

We propose a brief characterisation of the role played by each group, using the information provided by this connection probability matrix, in conjunction with Tables \ref{tab:degrees}, \ref{tab:confusion_table} and \ref{tab:counts}.
\begin{itemize}
 \item \textbf{Group 1}: this group contains inactive nodes only. These members do not send emails, but they may receive a few newsletters. 
 \item \textbf{Groups 2 to 8}: these contain a massive portion of the nodes, and hence describe the most common connectivity profiles that may arise.
 The nodes in these groups are not particularly active (the connection probabilities are typically not large) and tend to receive more emails than they send. 
 The sizes of these groups swell increasingly by acquiring nodes from group 1 until the bankruptcy, signalling increased activity in the months before the default.
 \item \textbf{Groups 9, 11, 12, 13}: these groups are mainly composed of managers, and correspond to active nodes, who send more than they receive. 
 These groups represent the different profiles of the executives of the company.
 Also note that these groups have a high within cluster connectivity, hence it is reasonable to find the corresponding conversations particularly meaningful in terms of intelligence and company directives.
 \item \textbf{Groups 10, 14, 15}: three groups of very active nodes, although these are mainly composed of employees. Note that groups 10 and 15 only become relevant during the year 2001.
 \item \textbf{Group 16}: this group actually only ever contains one person (with a non-specified role). This person apparently has some special position and hence may be considered as an outlier.
 \item \textbf{Group 17}: the smallest of groups, containing 4 unique members, each for $1$ time-frame only. 
 Since this group has high connection probabilities towards many groups, it is reasonable to believe that nodes join this group whenever they send out newletters to all members.
\end{itemize}

\begin{table}
\caption{Rounded average degrees for the Enron dataset, for each of the $17$ clusters found.}
\label{tab:degrees}
\centering
\begin{tabular}{lccccccccccccccccc}
  \toprule
 & 1 & 2 & 3 & 4 & 5 & 6 & 7 & 8 & 9 & 10 & 11 & 12 & 13 & 14 & 15 & 16 & 17 \\ 
  \midrule
  Avg. out-degree &   0 &   3 &   3 &   1 &   3 &   4 &   3 &   5 &   5 &   3 &   7 &  10 &  13 &  12 &  11 &  11 &  57 \\ 
  Avg. in-degree &   1 &   3 &   3 &   3 &   4 &   5 &   4 &   5 &   4 &   2 &   6 &   5 &   7 &   7 &   5 &   7 &   2 \\ 
  \bottomrule
\end{tabular}
\end{table}

\begin{table}
\caption{Cluster counts separated by status for the Enron dataset. For this table, the same node at two different time frames is considered as two separate entities, hence the sum of all entries is $TN$.}
\label{tab:confusion_table}
\centering
\begin{tabular}{lccccccccccccccccc}
  \toprule
 & 1 & 2 & 3 & 4 & 5 & 6 & 7 & 8 & 9 & 10 & 11 & 12 & 13 & 14 & 15 & 16 & 17 \\ 
  \midrule
  Employee & 576 &  90 &  36 &  45 &  92 &  50 &  61 &  22 &   7 &  46 &   1 &  18 &   0 &  17 &  18 &   0 &   1 \\ 
  Manager & 1027 & 231 & 191 & 102 &  34 &  83 &  53 &  16 &  89 &  40 &  87 &  45 &  41 &   6 &   5 &   0 &   2 \\ 
  N/A & 475 &  57 &  27 &  50 &  59 &  40 &  12 &  84 &   2 &   9 &   4 &   5 &   2 &  16 &   5 &  16 &   1 \\ 
  \bottomrule
\end{tabular}
\end{table}

\begin{table}
\caption{Group sizes at each time step for the Enron dataset.}
\label{tab:counts}
\centering
\footnotesize
\begin{tabular}{lccccccccccccccccc}
  \toprule
 Date & 1 & 2 & 3 & 4 & 5 & 6 & 7 & 8 & 9 & 10 & 11 & 12 & 13 & 14 & 15 & 16 & 17 \\ 
  \midrule
  2000-01-01 & 115 &   8 &   3 &   2 &   1 &   2 &   5 &   4 &   2 &   0 &   2 &   0 &   1 &   2 &   0 &   1 &   0 \\ 
  2000-02-01 & 114 &   5 &   6 &   2 &   1 &   2 &   5 &   4 &   2 &   0 &   3 &   1 &   0 &   2 &   0 &   1 &   0 \\ 
  2000-03-01 & 112 &   5 &   7 &   3 &   2 &   2 &   4 &   4 &   2 &   0 &   3 &   1 &   0 &   2 &   0 &   1 &   0 \\ 
  2000-04-01 & 107 &   5 &   8 &   4 &   4 &   3 &   4 &   3 &   2 &   0 &   3 &   1 &   0 &   3 &   0 &   1 &   0 \\ 
  2000-05-01 & 103 &   7 &   6 &   6 &   5 &   3 &   4 &   2 &   2 &   0 &   3 &   2 &   1 &   3 &   0 &   1 &   0 \\ 
  2000-06-01 &  96 &  10 &   7 &   6 &   6 &   3 &   5 &   3 &   2 &   1 &   3 &   2 &   1 &   2 &   0 &   1 &   0 \\ 
  2000-07-01 &  93 &  10 &   6 &   5 &   5 &   4 &   5 &   3 &   3 &   1 &   5 &   3 &   1 &   2 &   1 &   1 &   0 \\ 
  2000-08-01 &  87 &  14 &   4 &   5 &   6 &   5 &   4 &   4 &   3 &   2 &   6 &   3 &   2 &   2 &   0 &   1 &   0 \\ 
  2000-09-01 &  86 &  13 &   5 &   6 &   7 &   4 &   4 &   5 &   2 &   2 &   5 &   3 &   2 &   2 &   1 &   1 &   0 \\ 
  2000-10-01 &  79 &  16 &   7 &   5 &   8 &   5 &   5 &   5 &   3 &   0 &   5 &   4 &   2 &   2 &   1 &   1 &   0 \\ 
  2000-11-01 &  71 &  15 &   7 &   5 &   8 &   5 &   5 &   5 &   6 &   6 &   4 &   4 &   2 &   2 &   2 &   1 &   0 \\ 
  2000-12-01 &  71 &  14 &   6 &   7 &   9 &   5 &   6 &   5 &   6 &   5 &   5 &   3 &   1 &   2 &   2 &   1 &   0 \\ 
  2001-01-01 &  72 &  14 &   8 &   7 &   9 &   4 &   7 &   5 &   3 &   4 &   3 &   4 &   2 &   2 &   3 &   1 &   0 \\ 
  2001-02-01 &  68 &  17 &   7 &   8 &  10 &   5 &   7 &   5 &   3 &   4 &   3 &   3 &   2 &   2 &   3 &   1 &   0 \\ 
  2001-03-01 &  69 &  15 &   7 &   9 &  11 &   5 &   5 &   6 &   3 &   4 &   3 &   3 &   3 &   1 &   3 &   1 &   0 \\ 
  2001-04-01 &  61 &  18 &  10 &   8 &  10 &   6 &   4 &   6 &   4 &   6 &   3 &   4 &   4 &   1 &   2 &   1 &   0 \\ 
  2001-05-01 &  52 &  21 &  15 &   8 &  11 &   7 &   4 &   7 &   3 &   6 &   3 &   4 &   4 &   0 &   2 &   0 &   1 \\ 
  2001-06-01 &  58 &  17 &  19 &  10 &   9 &   7 &   4 &   7 &   3 &   6 &   4 &   1 &   1 &   1 &   1 &   0 &   0 \\ 
  2001-07-01 &  67 &  13 &  13 &   9 &   6 &   8 &   4 &   5 &   5 &   7 &   4 &   3 &   2 &   1 &   1 &   0 &   0 \\ 
  2001-08-01 &  58 &  15 &  13 &  10 &   9 &   9 &   4 &   5 &   5 &   7 &   4 &   3 &   3 &   1 &   1 &   0 &   1 \\ 
  2001-09-01 &  50 &  19 &  16 &  10 &  10 &  11 &   5 &   5 &   5 &   7 &   4 &   4 &   1 &   0 &   1 &   0 &   0 \\ 
  2001-10-01 &  42 &  25 &  16 &  11 &  10 &  11 &   5 &   5 &   5 &   7 &   4 &   4 &   1 &   0 &   1 &   0 &   1 \\ 
  2001-11-01 &  41 &  26 &  17 &  11 &  10 &  11 &   5 &   5 &   6 &   6 &   3 &   4 &   1 &   1 &   1 &   0 &   0 \\ 
  2001-12-01 &  56 &  20 &  17 &  12 &   6 &  11 &   4 &   4 &   6 &   4 &   3 &   2 &   1 &   1 &   1 &   0 &   0 \\ 
  2002-01-01 &  64 &  20 &  14 &  10 &   3 &  11 &   3 &   5 &   5 &   5 &   2 &   1 &   3 &   1 &   1 &   0 &   0 \\ 
  2002-02-01 &  78 &  13 &   9 &   9 &   5 &  11 &   5 &   2 &   5 &   5 &   1 &   1 &   2 &   1 &   0 &   0 &   1 \\ 
  2002-03-01 & 108 &   3 &   1 &   9 &   4 &  13 &   4 &   3 &   2 &   0 &   1 &   0 &   0 &   0 &   0 &   0 &   0 \\ 
  \bottomrule
\end{tabular}
\end{table}

\begin{figure}
 \centering
 \includegraphics[width=0.49\textwidth]{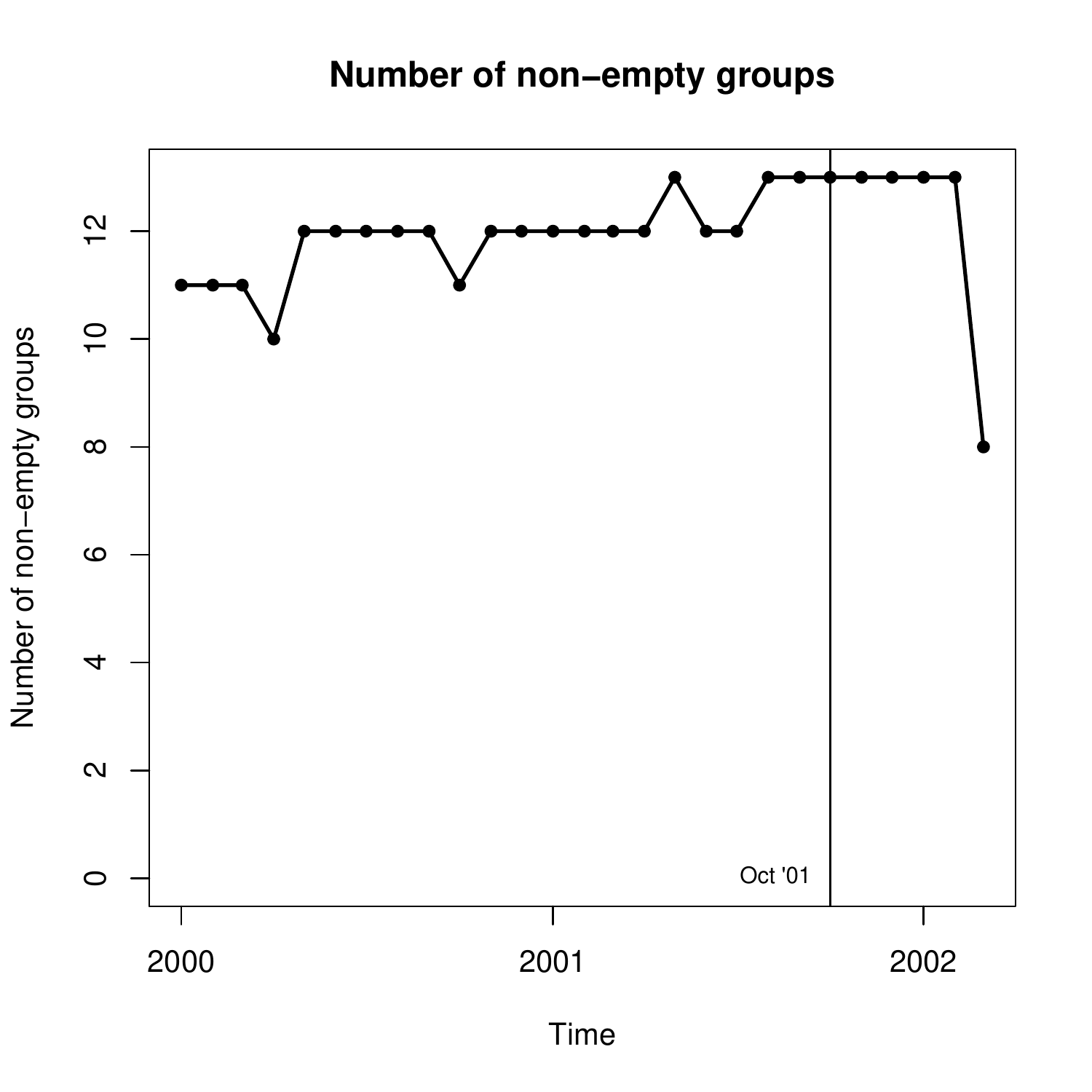}
 \includegraphics[width=0.49\textwidth]{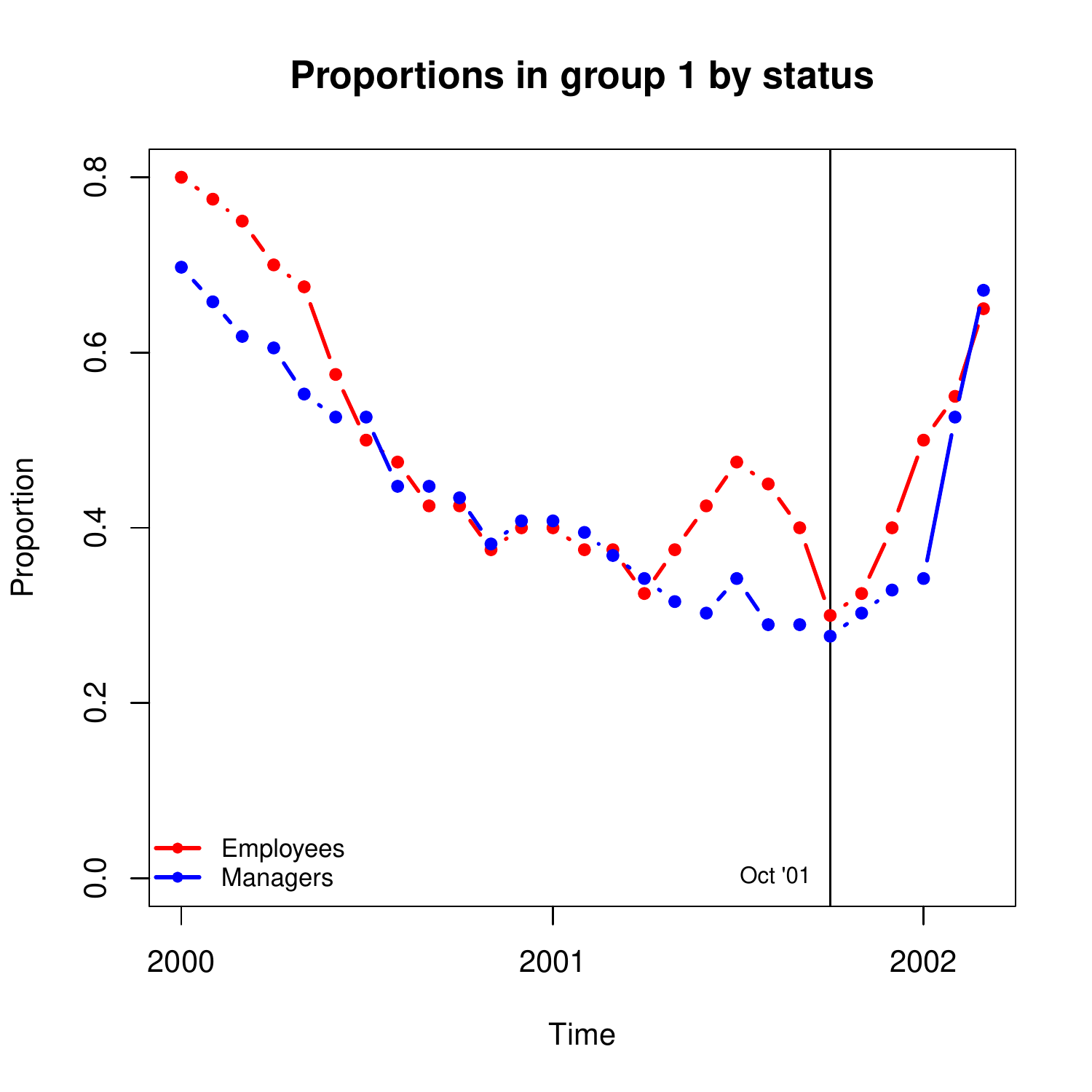}
 \caption{\textbf{Enron dataset}. On the left panel, the evolution of the number of non-empty groups is shown. This quantity appears to increase until the collapse, signalling an increase in heterogeneity in the network.
 On the right panel, the proportions of managers and employees allocated to the first group are shown. These proportions correspond to the nodes that are not particularly active.}
 \label{fig:enron_heterogeneity}
\end{figure}

\paragraph{Temporal dynamics.}
Our model allows for groups to become \textit{inactive}, whenever their sizes decrease to zero. 
This means that the connection profile associated to a particular group is not exhibited by any of the nodes in the time frame considered.
Hence, the number of non-empty groups can be used as a measure of heterogeneity in the network, and the temporal dynamics of this quantity can be used to assess how heterogeneity evolves over time.
In the left panel of Figure \ref{fig:enron_heterogeneity}, the evolution of the number of non-empty groups is shown. 
It appears that the network is particularly heterogeneous in the year before the collapse ($K^{(t)} = 15$ or $16$), whereas it eventually becomes rather homogeneous ($K^{(t)} = 10$) after the default.
A similar message is conveyed by the plot on the right panel of Figure \ref{fig:enron_heterogeneity}. 
Here the number of inactive nodes (i.e. nodes allocated to group 1) is shown to decrease convincingly for both employees and managers.
Eventually, after the collapse, the first group is repopulated as members quit their jobs and activities.

% \paragraph{Stability of managers and employees.}
% As shown on the left panel of Figure \ref{fig:swaps}, the managers appear to be more instable than employees: a few of them end up changing their allocation in (up to) $13$ out of $27$ time frames.
% \begin{figure}
%  \centering
%  \includegraphics[width=0.49\textwidth]{enron/swaps_per_node_by_status.pdf}
%  \includegraphics[width=0.49\textwidth]{enron/swaps.pdf}
%  \caption{Enron dataset. On the left panel, nodes are characterised based on the number of times they change group memberships, for both managers and employees. Some of the managers appear to change their allocations quite often.
%  On the right panel, the proportion of nodes changing group membership is plotted against time, for both managers and employees. 
%  Both classes become increasingly unstable, underlining a marked difference between the initial time frames and the remaining period.}
%  \label{fig:swaps}
% \end{figure}
% On the right panel of the same figure the proportion of managers and employees changing their allocation is plotted against time. 
% Both classes become increasingly unstable throughout the study, with peaks just before and right after the default.
% The increased instability before the default may be interpreted as a signal for the forthcoming disclosure, whereas the peaks after the default are due to nodes moving back to group 1.

\section{London bike sharing}\label{sec:london}
Data obtained from bike sharing systems is particularly suited to perform network analysis. 
Statistical analysis of the flows of bikes may provide important information regarding the management of the system and could help increase the efficiency of the service.
Cycle hires data can be easily rearranged and visualised as a dynamic network structure, where edges correspond to hires and nodes to stations. 
Similar approaches have been recently proposed in a number of works, such as: 
\textcite{guigoures2015discovering}, \textcite{randriamanamihaga2014clustering} and \textcite{matias01245867}.
In a similar fashion, here we propose an application of our methodology to a dataset of cycle hires in London. 

\begin{figure}
 \centering
 \includegraphics[width=0.49\textwidth]{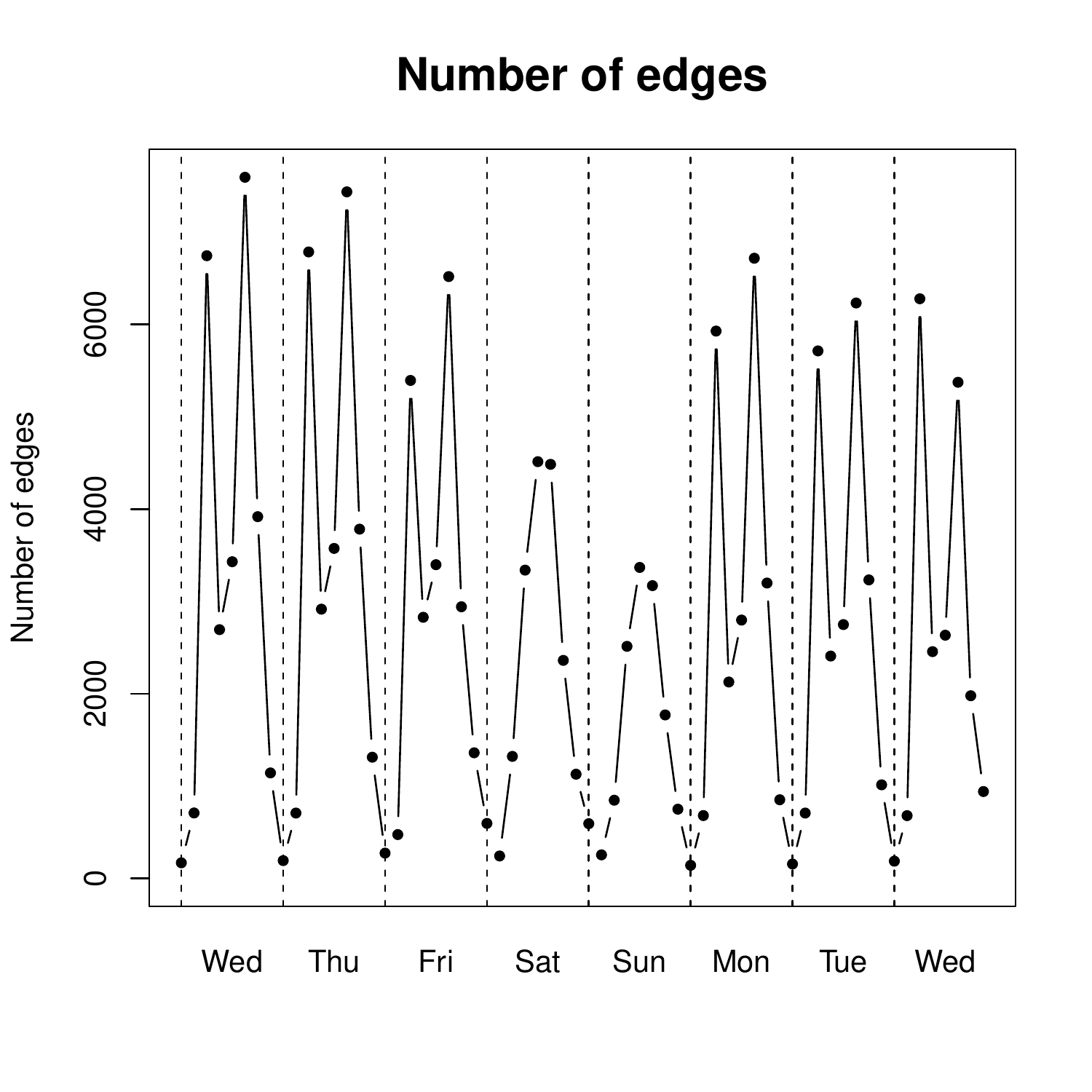}
 \includegraphics[width=0.49\textwidth]{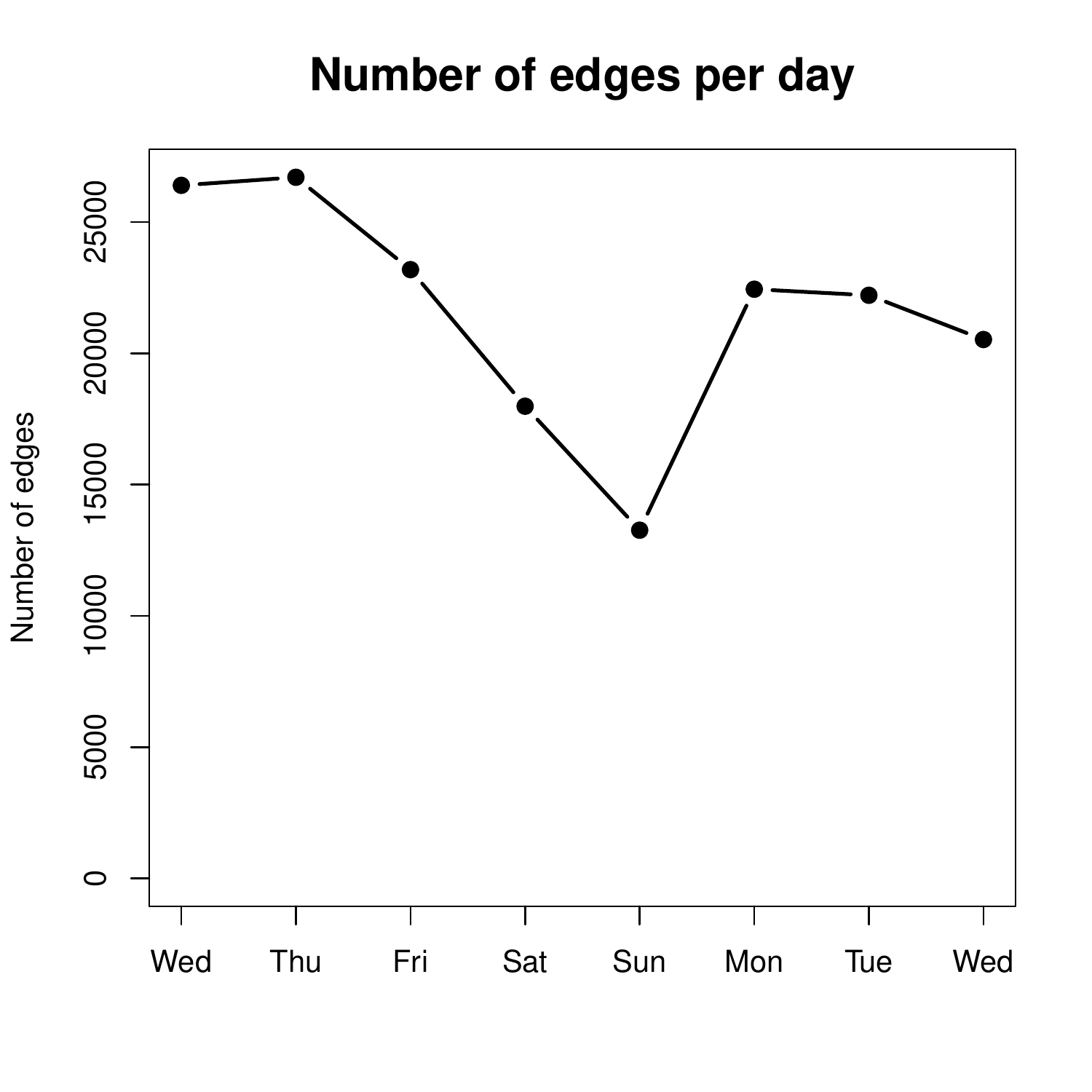}
 \caption{The number of edges in each time frame for the London bikes dataset are shown on the left panel. The right panel shows instead the total number of edges in each day.}
 \label{fig:n_edges}
\end{figure}

\paragraph{The data.}
The cycle hire data for the London bike sharing system is publicly available from \textcite{transport2016}. 
We use here the data from Wednesday, 5 June 2013 to Wednesday, 12 June 2013 (included). 
The discrete time frames correspond to blocks of three hours.
For each time frame, we create a network adjacency matrix by transmuting bike trips into directed edges: 
an edge from station $i$ to station $j$ appears at time frame $t$ if at least one bike is hired at station $i$ during the corresponding three hours, 
and the same bike is then returned to station $j$. 
Bikes may be returned to the same station where they were hired: we consider this information not important for our analysis and hence discard all of the self-edges. 
The dynamic network so obtained is made of $N=566$ nodes and $T=64$ time frames.
The data exhibits a very strong temporal heterogeneity, mainly due to the day-night cycle and the presence of the weekend days.
In Figure \ref{fig:n_edges} the observed number of edges is shown at every time frame. 
Weekdays typically exhibit two peaks in the activity level, corresponding to the start and the end of office hours. This suggests that commuters are the main users during working days.
By contrast, low activity is observed throughout the weekend, with a single activity peak appearing in the late mornings.

\paragraph{Results. }
We run our algorithm \texttt{GreedyIcl} once for each of the initialisation methods, with each run taking about six hours long. 
The optimal clustering found has a total of $K=43$ groups. 

\begin{figure}
 \centering
 \includegraphics[width=0.49\textwidth]{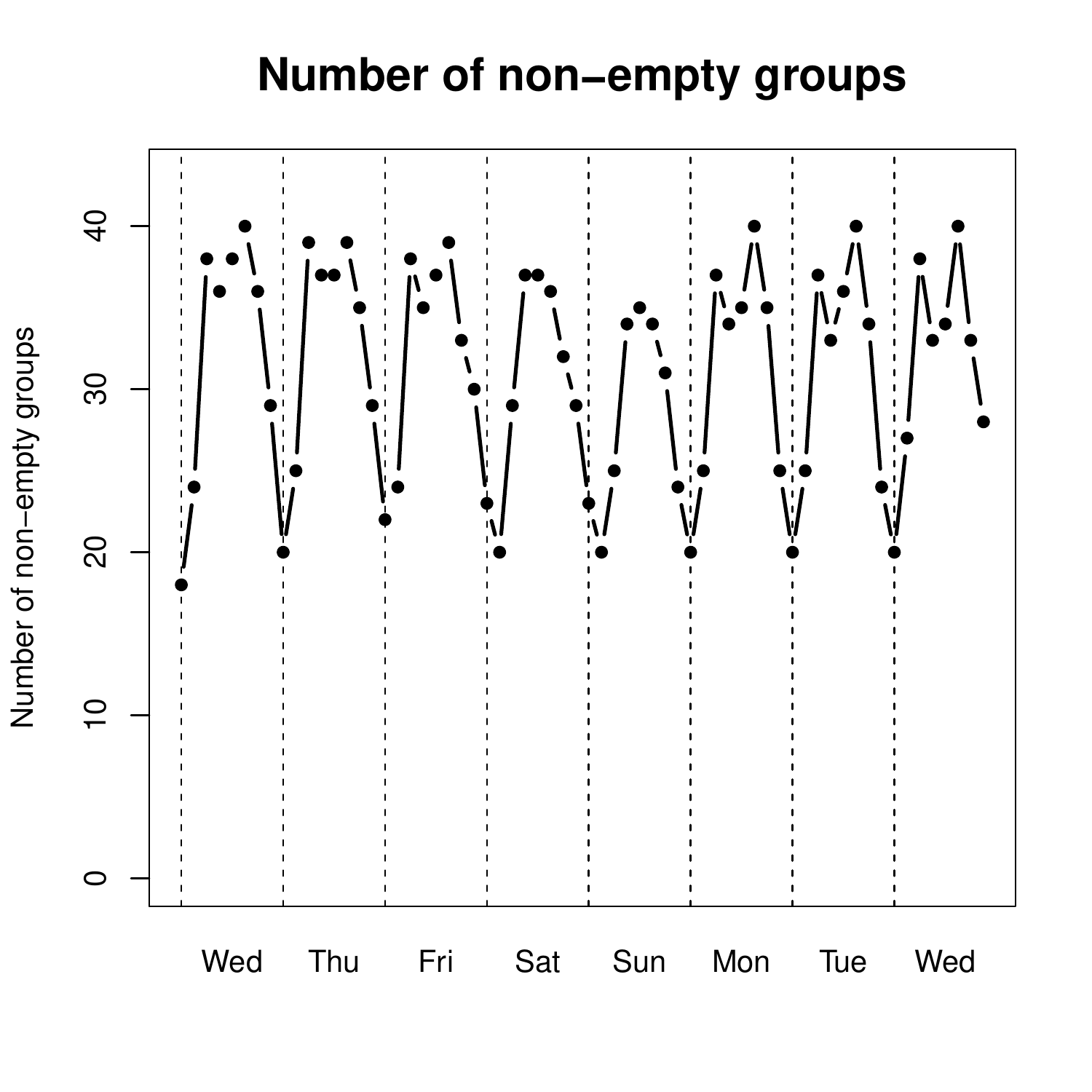}
 \caption{This plot shows the number of non-empty groups at each time frame for the London bikes dataset. The total number of unique groups found is $43$.
 The level of heterogeneity in the network has a noticeable temporal dynamic, which mimics the evolution of the number of links shown in Figure \ref{fig:n_edges}.}
 \label{fig:heterogeneity}
\end{figure}

\paragraph{Heterogeneity. }
The temporal dynamic of the activity level is also exhibited in Figure \ref{fig:heterogeneity} by the number of active groups at each time frame: 
as in the Enron dataset, we use this as a measure for the level of heterogeneity in the network.
It appears that, as activity peaks in weekdays, the number of active groups doubles, signaling a heavy increase in heterogeneity within the network.
Surprisingly, weekend days do not exhibit a markedly different heterogeneity level with respect to weekdays, even though the overall activity level is lower.
This suggests that fewer nodes become active in the weekend, but their connection patterns are not particularly different than those seen in weekdays.

\paragraph{Characterisation of the groups.}
The groups can be roughly divided in two categories: 
the first $20$ groups have a large aggregated size, mostly high stability, low connection probabilities and a good balance between expected out-degree and in-degree.
The rest of the groups have smaller sizes, high instability, higher connection probabilities and in some cases very unbalanced out-degrees and in-degrees. 

\begin{figure}
 \centering
 \includegraphics[width=0.32\textwidth,page=2]{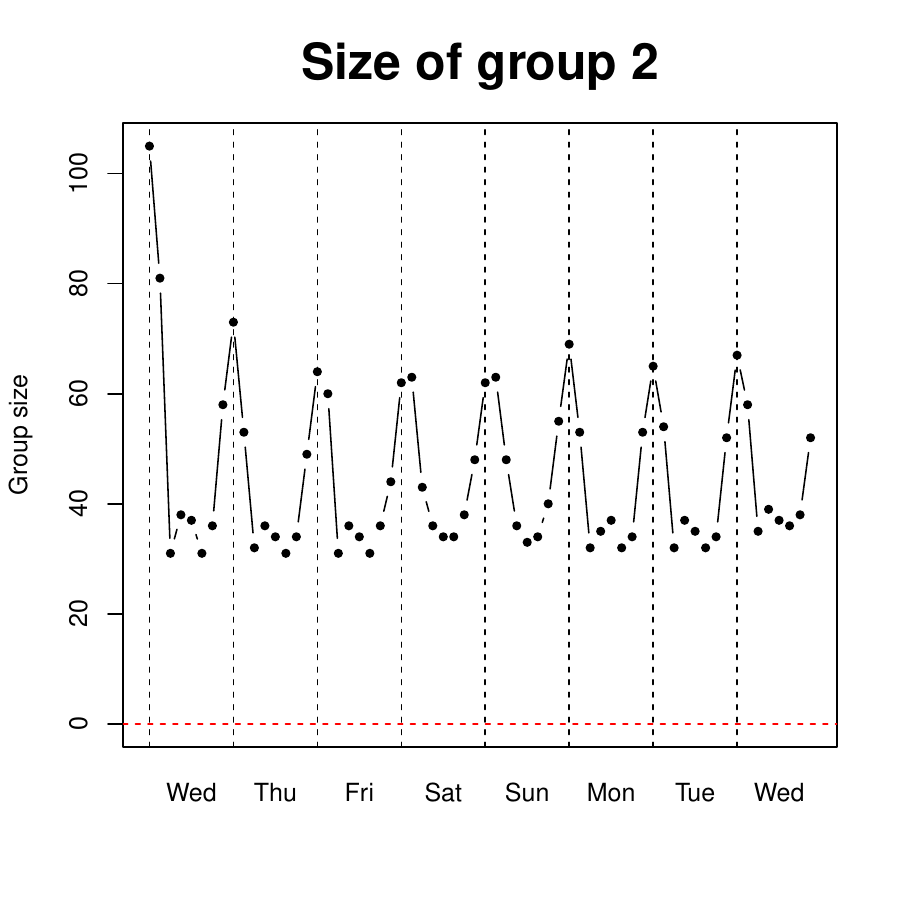}
 \includegraphics[width=0.32\textwidth,page=6]{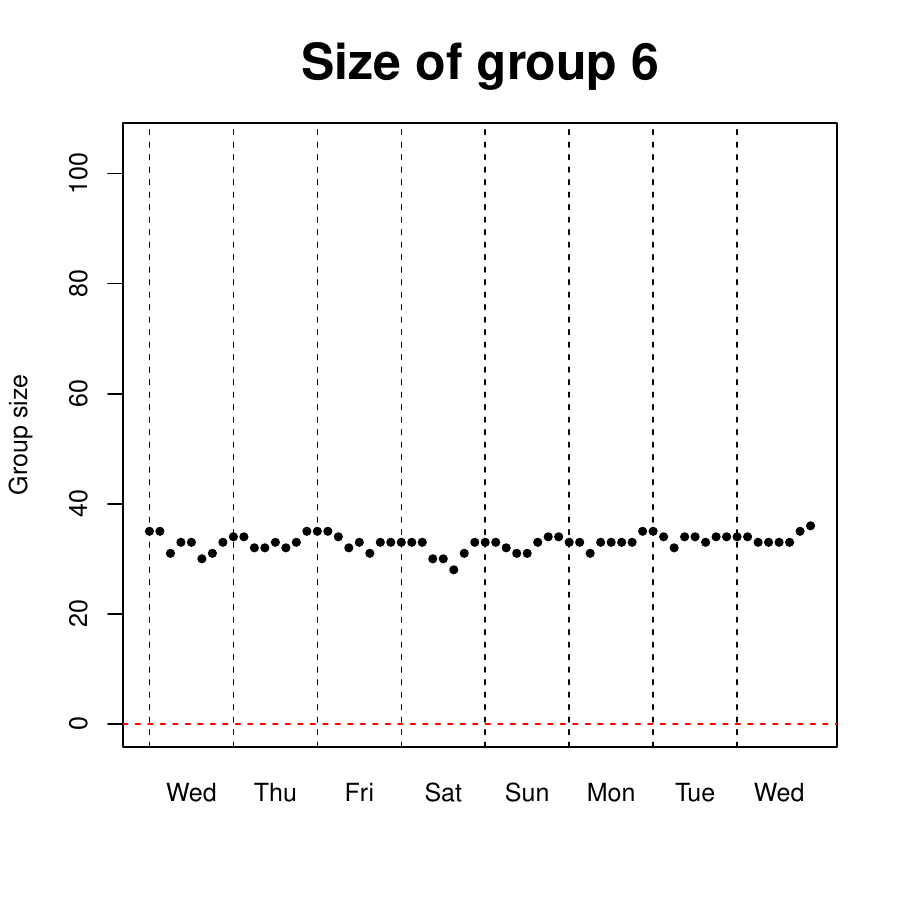}
 \includegraphics[width=0.32\textwidth,page=12]{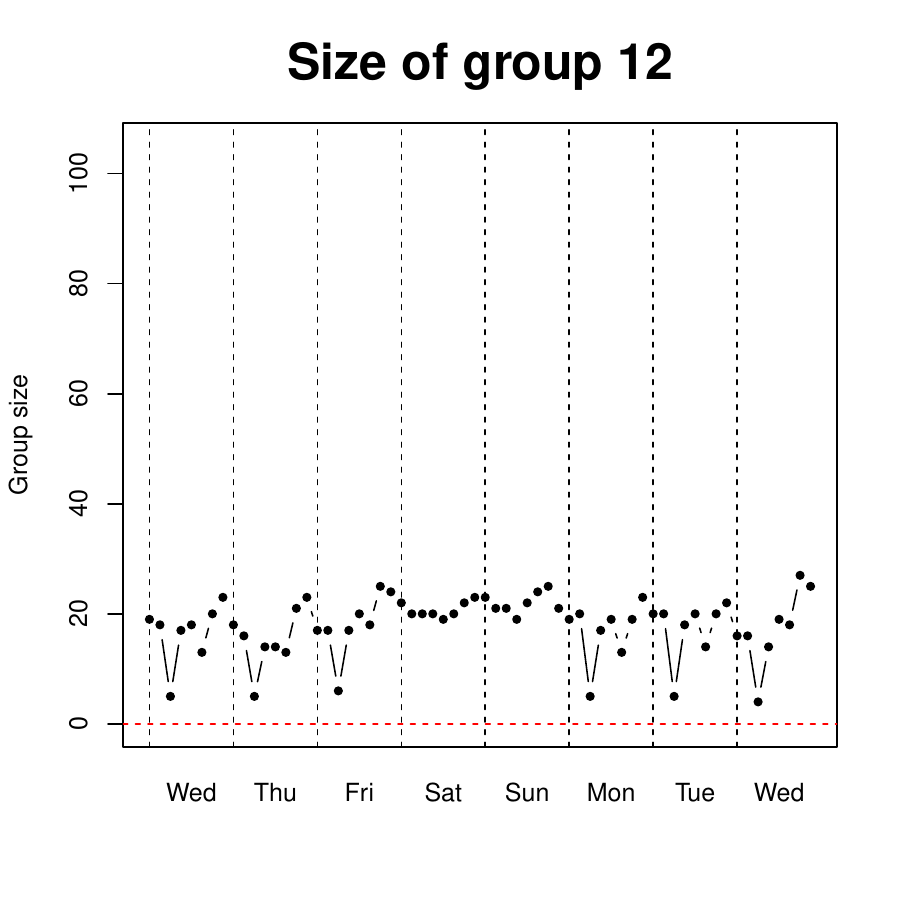}
 \includegraphics[width=0.32\textwidth,page=18]{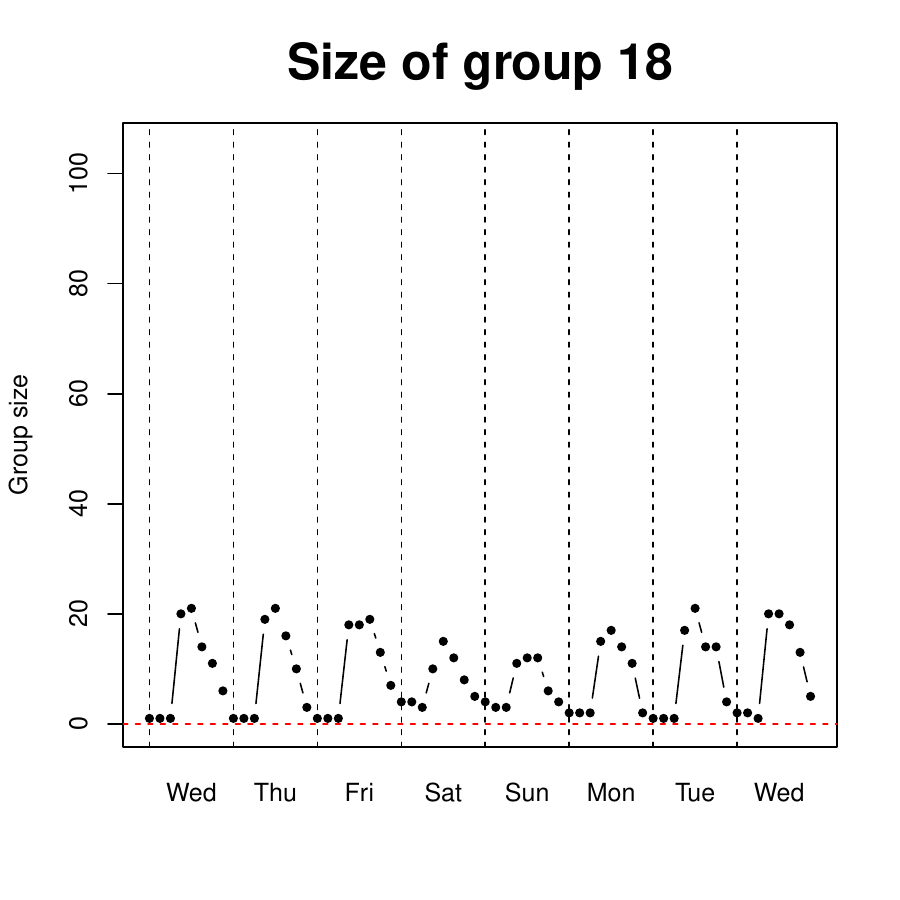}
 \includegraphics[width=0.32\textwidth,page=21]{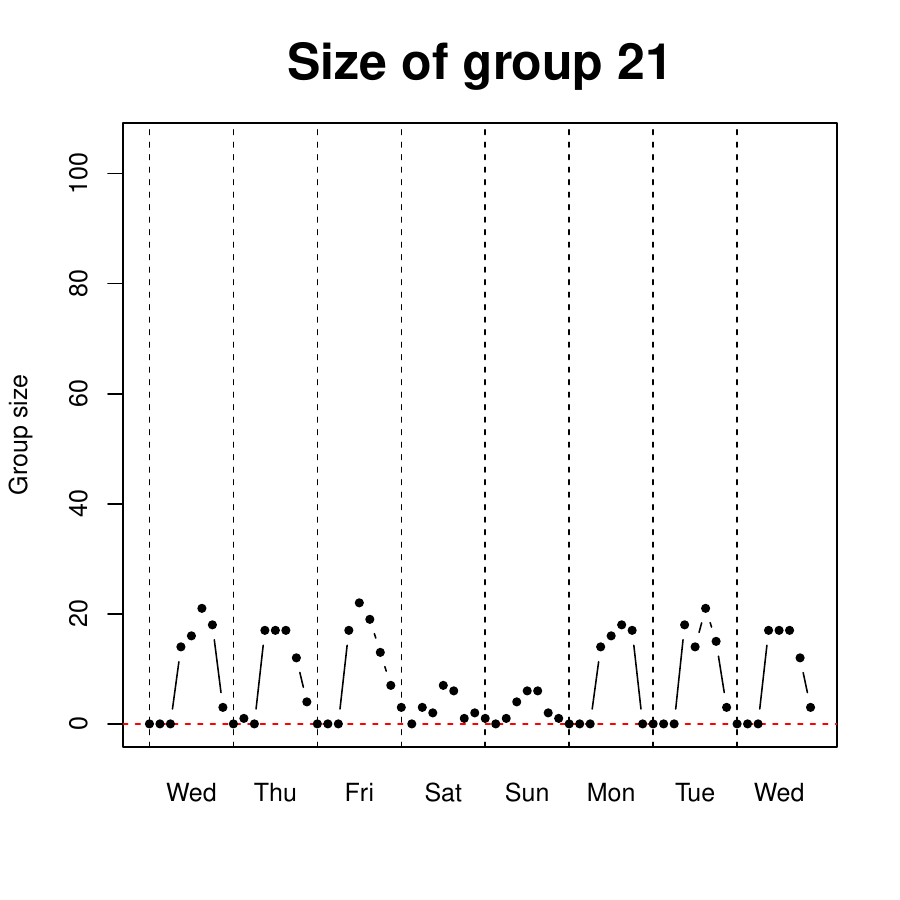}
 \includegraphics[width=0.32\textwidth,page=33]{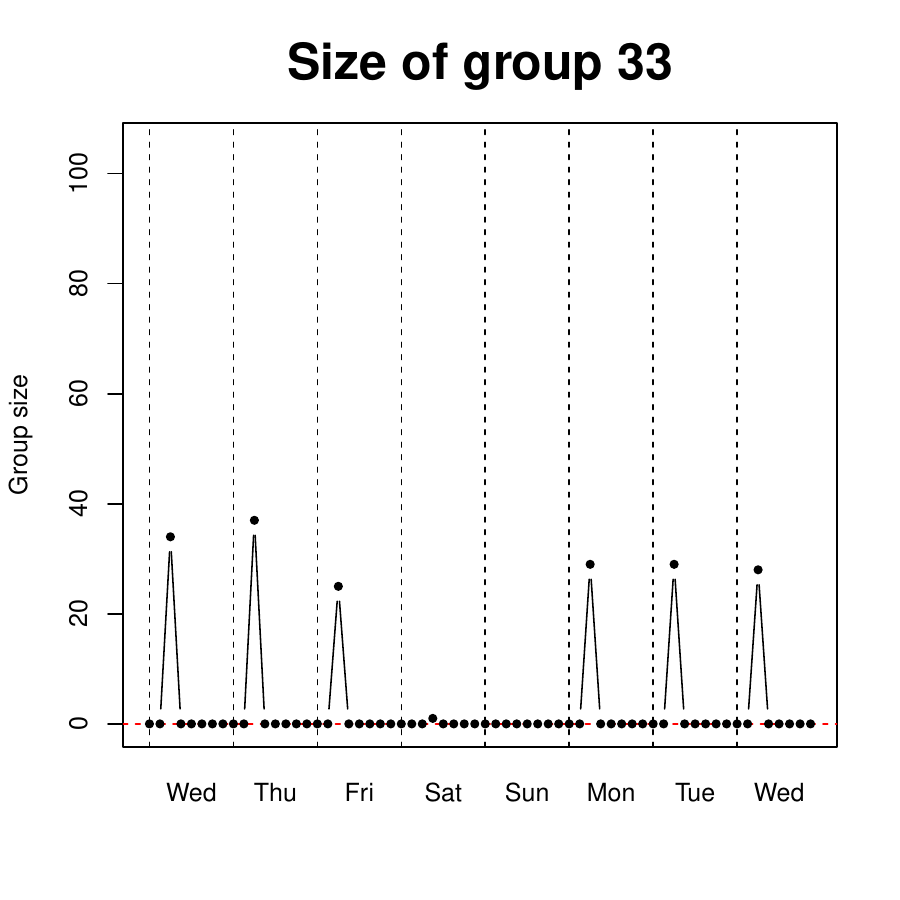}
 \caption{\textbf{London bikes data}. Temporal dynamics for the sizes of some of the groups found. 
 The three groups on the upper row belong to the first category (groups containing mostly inactive nodes): group 2 swells consistently every night; group 6 is stably present throughout; whereas group 12 is almost emptied in every weekday at the start of the office hours.
 The groups on the lower row contain instead very active stations: group 18 is emptied every night; group 21 is also emptied throughout the weekends; whereas group 33 activates every weekday at the start of office hours. }
 \label{fig:counts}
\end{figure}

Figure \ref{fig:counts} shows the temporal evolution of the size of a selection of groups. 
It appears that the trends in the activity level have a huge effect on the migrations of stations between groups. 
As nodes become more active, they leave the large groups (first category) and move to smaller groups, which are better suited to capture the details of their new connection profile in the network.
During the high-activity regime these nodes may end up visiting several groups based on the evolution of their connectivity patterns.
Nonetheless a good portion of nodes is not affected by the variation of activity level and simply remains stably in the larger and inactive groups.

\begin{figure}
 \centering
 \includegraphics[width=0.49\textwidth]{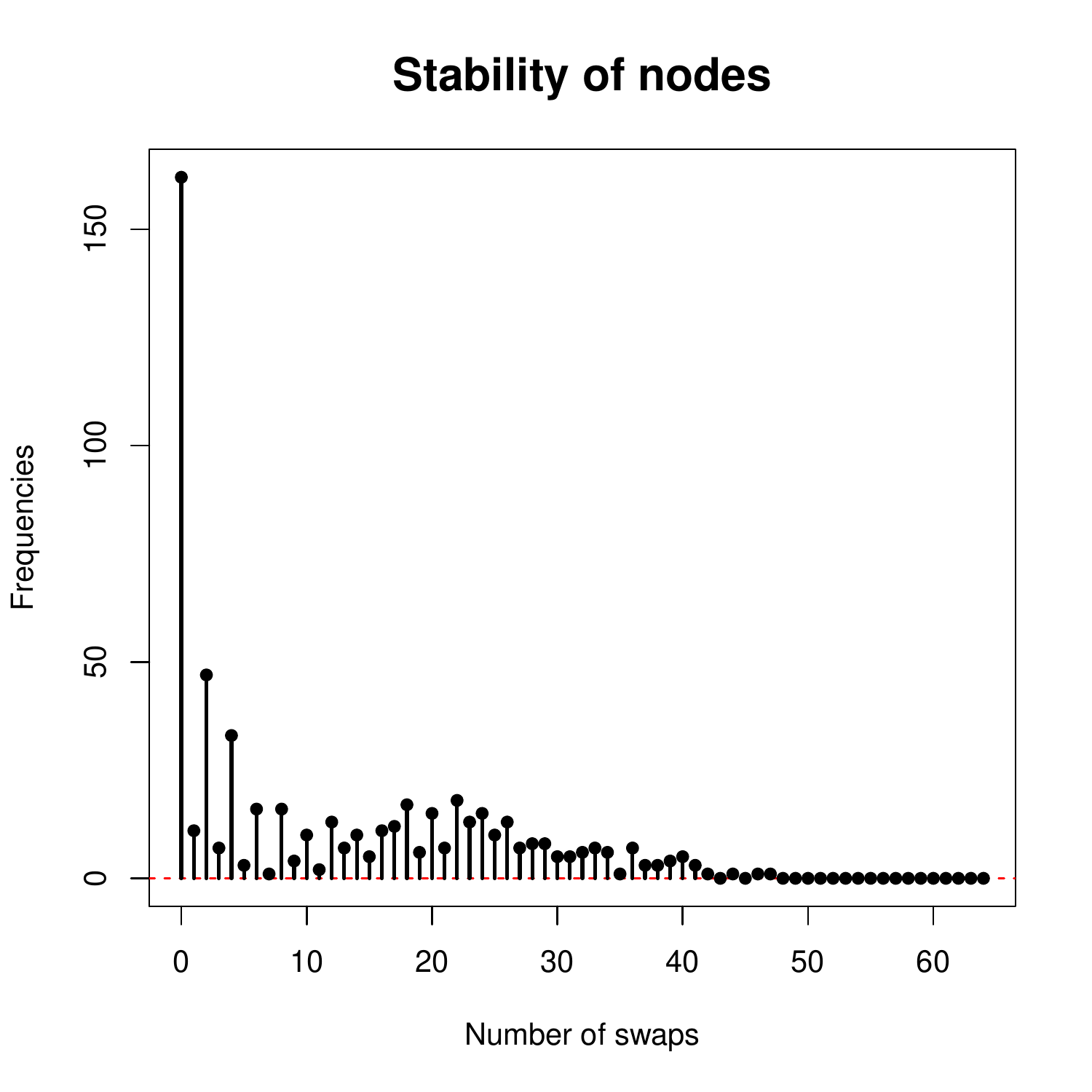}
 \includegraphics[width=0.49\textwidth]{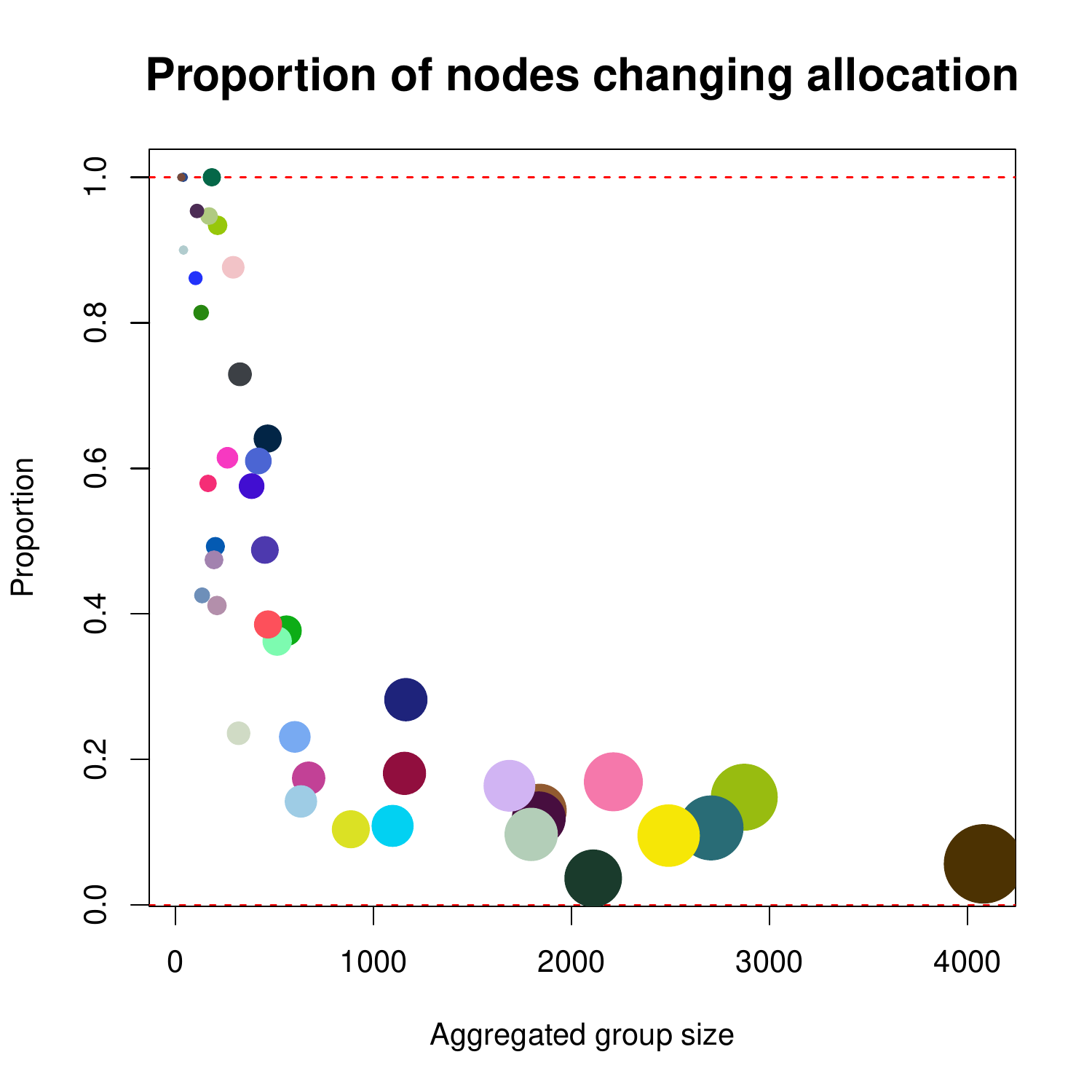}
 \caption{Stability level in the London bikes dataset. The left panel shows that although a good portion of nodes are very stable, there are some stations changing their allocations in more than 45 out of 64 time frames. 
 The plot on the right hand side studies instead the stability of groups. It appears that larger groups are very stable, and viceversa smaller groups are not. The size of the circles is proportional to the size of the group aggregated over time.}
 \label{fig:stability1}
\end{figure}

These arguments are very well supported by the plots in Figure \ref{fig:stability1}.
It appears that the stability of nodes is very heterogeneous in that there are many very stable nodes but also some very unstable ones (left panel).
At the same time, the plot on the right hand side reinforces the idea that larger groups are also very stable whereas smaller groups exhibits high instability.

\begin{figure}
 \centering
 \includegraphics[width=0.49\textwidth]{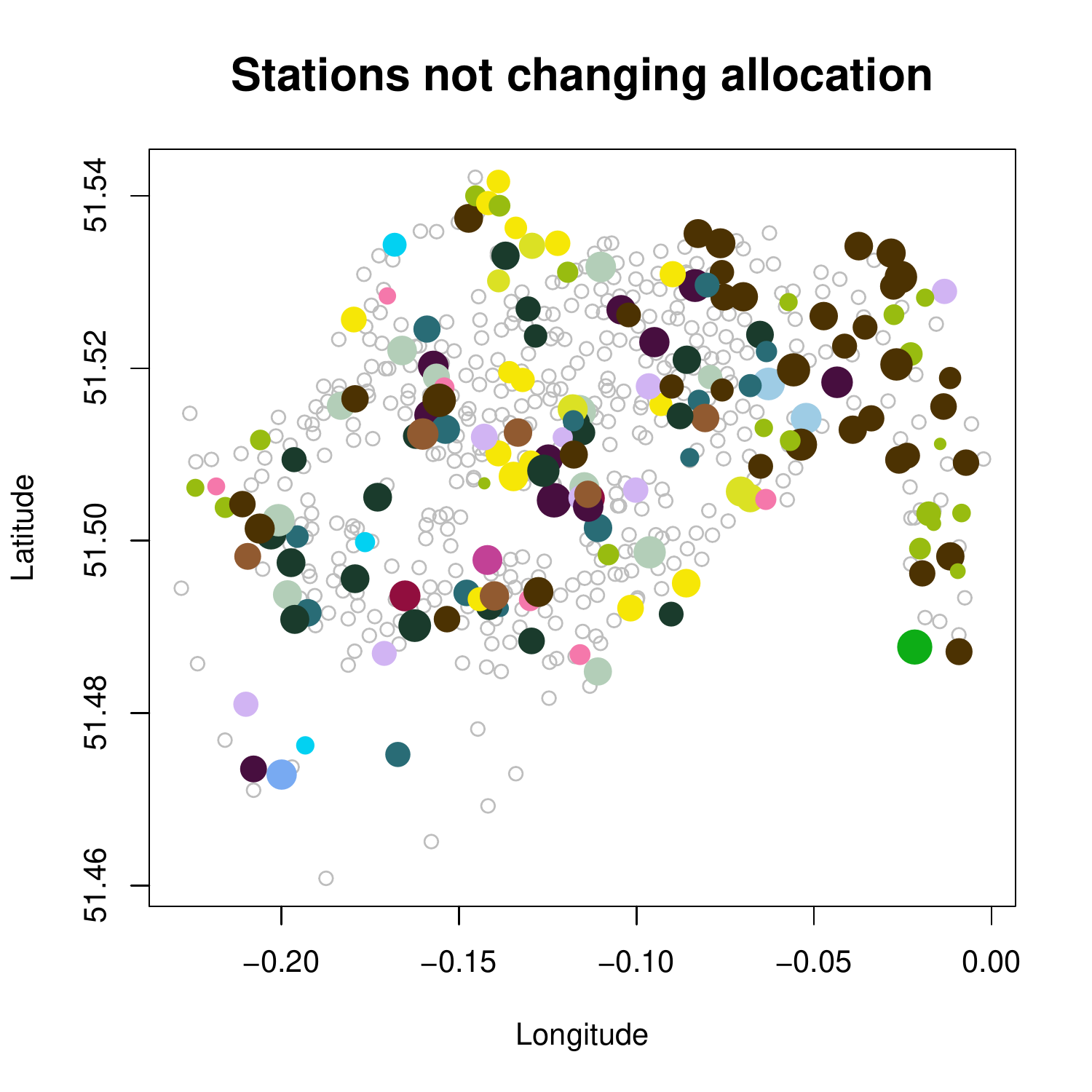}
 \includegraphics[width=0.49\textwidth]{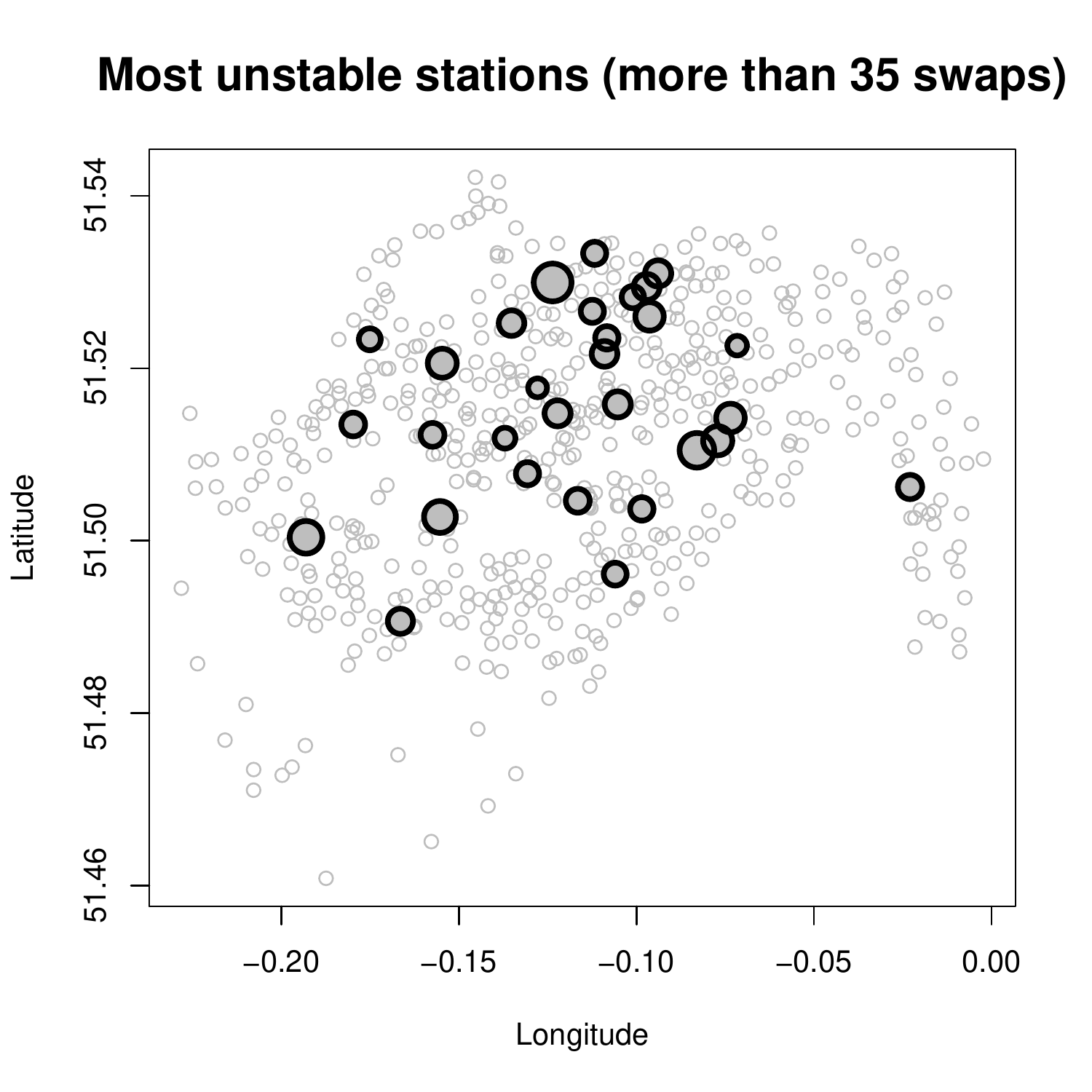}
 \caption{On the left panel, the spatial distribution of the stations not changing allocations is shown for the London bikes dataset. The colour of the circles correspond to the group they are allocated to at every time frame. 
 On the right panel, instead, the stations performing more than 35 swaps are shown. In both plots the size corresponds to a combination of the out-degree and in-degree of the nodes.}
 \label{fig:stability2}
\end{figure}

As shown in Figure \ref{fig:stability2} the stability does not seem to be particularly related to the geographical position of stations.
Furthermore, the number of groups containing stable stations (nodes that never change allocation) is relatively high.

\begin{figure}
 \centering
 \includegraphics[width=0.49\textwidth]{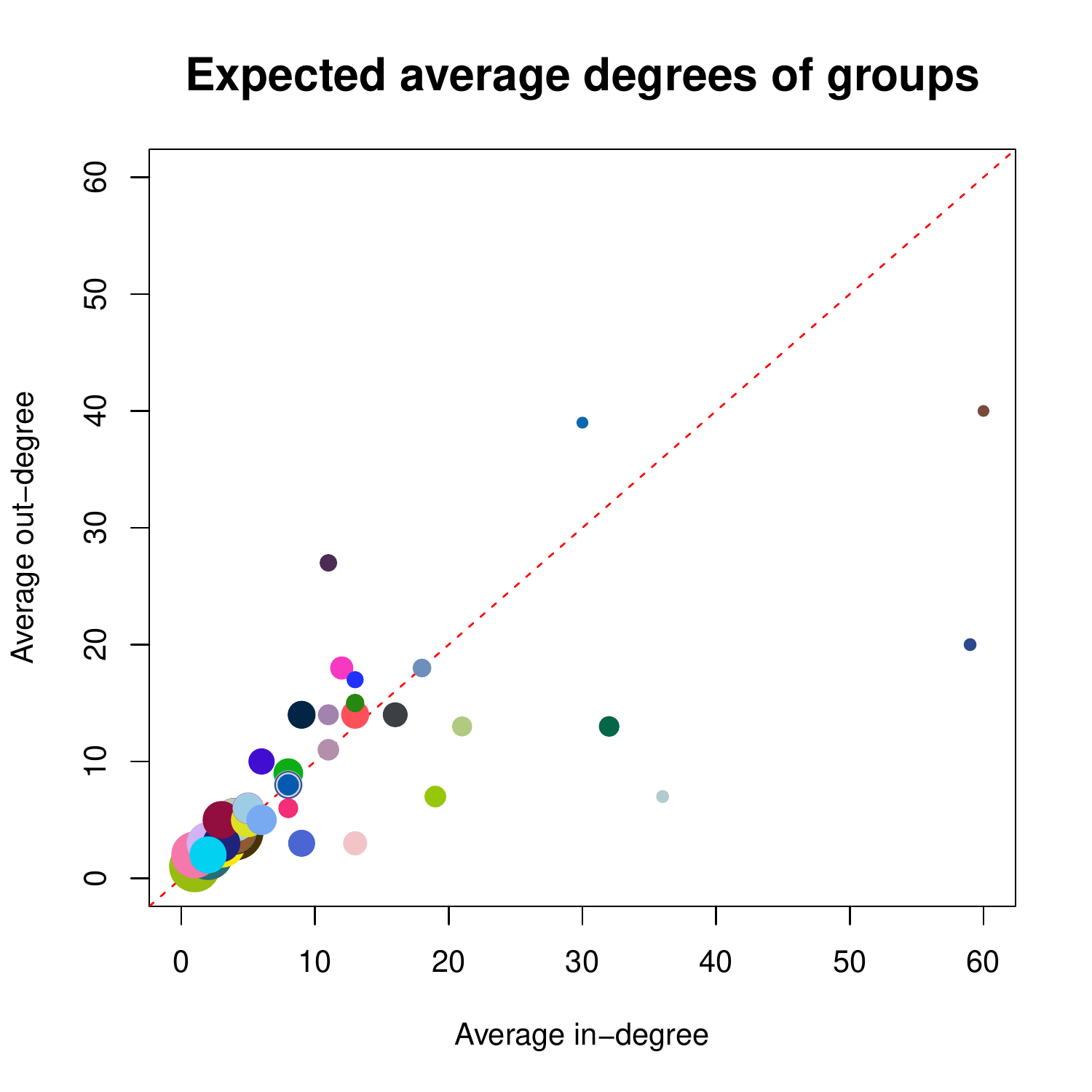}
 \caption{Expected out-degree versus expected in-degree for the groups found in the London bikes dataset. The size of circles is proportional to the size of groups aggregated over time. 
 All of the large groups of inactive nodes have balanced degrees, in that they tend to send and receive the same number of edges. 
 Smaller groups, instead, can exhibit very unbalanced degrees, either in favour of the out-degree or in-degree.}
 \label{fig:degrees}
\end{figure}

One additional feature that distinguishes the two categories of groups found can be observed in Figure \ref{fig:degrees}.
Here, for every group, the expected out-degrees versus the expected in-degrees are plotted. 
While stable groups (first category) tend to exhibit a good balance between the two degrees, smaller and unstable groups can also have very unbalanced degrees.
This means, evidently, that whenever a node joins one of this unbalanced groups, it may send many more edges than what it receives (or viceversa).
In this bike sharing context, the corresponding stations may require special attention due to temporary excess of hiring demands (or excess of arrivals).

\section{Conclusions}
The statistical analysis of dynamic networks is particularly challenging, both from a modelling point of view (due to the temporal dynamics) and 
from an estimation point of view, due to the inherent computational difficulties.
In this paper we have focused on an extension of the stochastic block model to dynamic networks, where the temporal evolution of the nodal information is captured by a Markovian process.
Our formulation allows one to integrate out all of the model parameters from both the likelihood and the prior, thereby obtaining an analytical formula for the marginal posterior of the allocation variables.
In a model-based clustering context, such a marginal posterior is equivalent to the exact integrated completed likelihood, which is widely used as an optimality criterion for partitions.

Taking advantage of these results, we have proposed a greedy algorithm to estimate the optimal clustering of the nodes. 
Our algorithm resembles other tools proposed in the recent literature, and scales particularly well with the size of the data. 
However, only convergence to a local optimum is guaranteed, hence several restarts and a careful initialisation are particularly useful in order to find the global optimum.

One appealing feature of our approach is that model-choice is carried out automatically, since the number of groups can be deduced at each time frame from the optimal clustering solution.

Through a simulation study, we have validated both our optimality criterion and the greedy algorithm used to estimate the optimal partition. 
Also, we have compared our method to one based on a variational Expectation-Maximisation algorithm, 
showing that with careful initialisation our approach achieves better results when the underlying number of groups is unknown.

We have applied our methodology to the Enron email dataset, and found $17$ underlying groups. 
Each of the groups found appears to have a specific role within the network and the original status of the members seem to be particularly related with the nodes' allocations.
Also, we have analysed some relevant summaries from our optimal clustering underlining the dynamic evolution of the heterogeneity within the network and of the overall activity level.

We have also applied our methodology to a London Bikes dataset, to study the flows and connections between bike stations. 
Our method has returned an optimal clustering made of $43$ groups. 
The analysis of such a complex structure has been particularly challenging but we have managed to extract some interesting information regarding the dynamics captured by the model.

We note that, while the practice of integrating out the likelihood parameters has been exploited in a number of recent papers, 
the collapsing of the transition probabilities in the Markov process is a rather new original technique. 
Not only this creates an ideal setup for our method, but it may also be generalised and exploited in arbitrary discrete hidden Markov models and their extensions.

Additionally, our work may be extended in a number of ways.
We have used our algorithm only on binary networks, but this approach generalises to networks with other edge types. 
Incomplete weighted graphs can be easily handled using a Bernoulli presence-absence indicator, and hence using the same framework based on the collapsing of the parameters.

The initialisation plays an important role in our method, yet finding a good starting partition is a great challenge. 
In fact, there is a lack of scalable methods that can handle complex dynamic objects such as networks. 
Furthermore, clustering cannot be attempted for each time frame independently since this result in label-switching problems.
Here, we have proposed two new intialisation methods which may in some cases improve the results.

Extensions to the supervised classification case are also straightforward: if some of the allocations are known, these need not be updated during the greedy optimisation.
From a Bayesian perspective, the optimal clustering solution obtained will then maximise the posterior predictive distribution, rather than the posterior. 
This strategy would be useful when dealing with nodes that join or leave the study dynamically, since, as suggested by \textcite{matias2015statistical}, a fixed group of inactive nodes may be created.

More generally, the optimisation of the exact Integrated Completed Likelihood is a particularly challenging task, due to the discrete search space and the pronounced multimodality of such objective function. 
The \texttt{GreedyIcl} algorithm introduced in this paper was shown to perform well, although future extensions may improve upon our work in terms of efficiency using alternative optimisation routines.

\section*{Acknowledgements}
The Insight Centre for Data Analytics is supported by Science Foundation Ireland under Grant Number SFI/12/RC/2289.
Riccardo Rastelli and Nial Friel's research was supported by a Science Foundation Ireland grant: 12/IP/1424.

% \nocite{*}
\printbibliography

\end{document}